\documentclass[11pt,a4paper]{article}

\usepackage{latex/usepackages}

\input{latex/latex-commands.sty}
\input{latex/physics.sty}

\title{
Bridging the divide: axion searches and axino phenomenology at colliders
}

\date{\today}

\author[a,b,c]{Gabe Hoshino,}
\author[a,b]{Kristin Dona,}
\author[a,b,c,d,e]{Keisuke Harigaya,}
\author[a,b,c]{David W. Miller,}
\author[a,b,*]{\newline Jan T. Offermann,%
  \note[*]{Now at Brown University.}}
\author[a, b,**]{Bianca Pol,%
  \note[**]{Now at University of California, Berkeley.}}
\author[a,b]{and Benjamin Rosser}

\affiliation[a]{Department of Physics, University of Chicago,\\ 933 East 56th Street, Chicago, IL 60637, U.S.A.}
\affiliation[b]{Enrico Fermi Institute, University of Chicago,\\ 5720 South Ellis Avenue, Chicago, IL 60637, U.S.A.}
\affiliation[c]{Kavli Institute for Cosmological Physics, University of Chicago,\\ 5640 South Ellis Avenue, Chicago, IL 60637, U.S.A.}
\affiliation[d]{Leinweber Institute for Theoretical Physics, University of Chicago,\\ 933 East 56th Street, Chicago, IL 60637, U.S.A.}
\affiliation[e]{Kavli Institute for Physics and Mathematics of the Universe (WPI), The University of Tokyo Institutes for Advanced Study, and The University of Tokyo,\\ 5-1-5, Kashiwanoha, Kashiwa, Chiba 277-8583, Japan}

\emailAdd{gyhoshino@uchicago.edu}
\emailAdd{kdona@uchicago.edu}
\emailAdd{kharigaya@uchicago.edu}
\emailAdd{David.W.Miller@uchicago.edu}
\emailAdd{jan\_offermann@brown.edu}
\emailAdd{bpol@berkeley.edu}
\emailAdd{brosser@uchicago.edu}

\abstract{We discuss a phenomenological model that extends the minimal supersymmetric standard model to contain axions and their supersymmetric partner, the axino. In the supersymmetric DFSZ axion model, the axino has tree level couplings to the higgs sector. In the case where $R$-parity is conserved, collider experiments may be sensitive to displaced decays of heavier neutralino states into lighter, mostly axino states. We present a sensitivity analysis using a model in which mostly higgsino next-to-lightest supersymmetric particle states decay into a mostly axino lightest supersymmetric particle. The model is studied using Monte Carlo simulation produced using \Madgraph and estimates of experimental sensitivities to the model, including detector simulation and kinematic selections, are evaluated using the \MadAnalysis{5} framework. For a higgsino mass below 1 TeV, the axion decay constant below $f_{a} < 10^{11}$ GeV can be effectively probed by the Large Hadron Collider with an integrated luminosity of 140 fb$^{-1}$. This work demonstrates that supersymmetric DFSZ axion models can be studied with existing collider experiments, offering complementary sensitivity to direct-detection and astrophysical searches and paving the way for broader exploration of supersymmetric axion scenarios.}


\arxivnumber{2511.07224}

\begin{document}

\maketitle

\section{Introduction}
\label{sec:intro}
The axion is a well-motivated dark matter candidate which was initially proposed by Peccei and Quinn to solve the strong-$CP$ problem dynamically~\cite{Peccei:1977hh,Peccei:1977ur}. The so-called $\theta$-term of QCD Lagrangian:
\begin{align}
\mathcal{L}_{\mathrm{QCD}} \supset \theta \frac{g^{2}}{32\pi^{2}}G \tilde{G},
\label{eq:theta-term}
\end{align}
together with non-perturbative QCD dynamics, explicitly violates $CP$-symmetry and would imply a nonzero neutron electric dipole moment (EDM)~\cite{tHooft:1976rip,Crewther:1979pi}. Measurements of the neutron EDM constrain it to be very small, $\theta < 10^{-10}$, which causes a fine-tuning problem in QCD~\cite{abel2020measurement}. To solve this problem dynamically, a new spontaneously broken global $\mathrm{U}(1)$ symmetry, Peccei-Quinn (PQ) symmetry, is introduced. The resultant pseudo-Nambu-Goldstone boson is called the axion. The $\mathrm{U}(1)_{\rm PQ}$ symmetry is anomalous and the strength of the coupling of the axion to the gluon is inversely proportional to the symmetry breaking scale, which is called the axion decay constant $f_a$. The axion obtains a potential by strong QCD dynamics and a vacuum expectation value (VEV) which cancels the $\theta$ term, thus solving the strong $CP$ problem dynamically.

Though the axion was introduced to solve the strong-$CP$ problem, the axion is also a natural dark matter candidate. The axion mass, $m_a$, and its couplings are inversely proportional to $f_a$, yielding a large range of $f_a$ for which the astrophysical lower bounds are satisfied (\cite{Caputo:2024oqc} and references therein) and the axion becomes long-lived and is coupled feebly-enough to particles of the standard  model (SM) to be the dark matter.
There are several known mechanisms to produce axions in the early universe to explain the observed dark matter abundance. In particular, the so-called misalignment mechanism predicts $f_a \lesssim 10^{12}$ GeV unless the initial condition of the axion field is fine-tuned~\cite{Preskill:1982cy,Abbott:1982af,Dine:1982ah}. In another mechanism, the observed matter-antimatter asymmetry of the Universe can also be explained~\cite{Co:2019jts,Co:2019wyp}.

Many prominent searches for axion dark matter rely on the axion-photon coupling.
Because of the electromagnetic anomaly of the $\mathrm{U}(1)_{\rm PQ}$ symmetry as well as the axion-meson mixing, the axion generically couples to the photon with a strength $g_{a\gamma\gamma} \propto 1/f_{a} \propto m_{a}$~\cite{raffelt1990astrophysical}. In a strong magnetic field, axions will convert into photons at some rate determined by the coupling $g_{a\gamma\gamma}$. Experiments sensitive to the conversion of dark matter axions into photons in a strong magnetic field can thus probe the $g_{a\gamma\gamma}$, $f_{a}$, and $m_{a}$ parameter space directly. In order to achieve sensitivity to the necessarily feeble axion-photon coupling, experiments must employ methods to significantly enhance the photon signal compared to noise and other backgrounds. The most prominent method is through resonant cavity enhancement like ADMX~\cite{ADMX:2018gho} or HAYSTAC~\cite{HAYSTAC:2018rwy} in the $\sim$$\mu \mathrm{eV}$ axion mass regime. Dielectric haloscopes like MADMAX~\cite{Li_2020} and plasma haloscopes like ALPHA~\cite{Millar_2023} seek to push sensitivity to the higher end of the $\sim$$\mu \mathrm{eV}$ range. For axion masses much higher than the $\sim$$\mu \mathrm{eV}$ scale, experiments like BRASS~\cite{bajjali2023first} and BREAD~\cite{liu2022broadband} use reflectors to focus an axion-photon signal onto a suitably low-noise photosensor. Finally, for much lower masses than resonant cavity haloscopes can probe, experiments like ABRACADABRA~\cite{Salemi:2021gck} and DM Radio~\cite{silva-feaver2017} use tunable LC circuits to detect an axion-photon signal. These experiments are relatively model independent as they search for a generic pseudoscalar dark matter particle with a non-zero photon coupling. These generic probes of pseudoscalar dark matter beyond the SM (BSM) can then be used to constrain specific UV-complete models such as the prominent Kim-Shifman-Vainshtein-Zakharov~\cite{kim_weak-interaction_1979, shifman_can_1980} (KSVZ) and Dine-Fischler-Srednicki-Zhitnitsky~\cite{zhitnitsky_possible_1980, dine_simple_1981} (DFSZ) QCD axion models, which have different coupling strengths to photons. As we will discuss, the introduction of further BSM physics such as supersymmetry can actually suppress the photon coupling, making these essential probes of axion dark matter significantly less sensitive to some well-motivated models.

There exists another issue of fine-tuning in the SM known as the electroweak hierarchy problem.
The mass of the Higgs boson is not protected by any symmetry, and as a result, the Higgs boson mass should receive quantum corrections from every massive particle in the SM, thus yielding a mass close to the Planck scale ($M_{\mathrm{Pl}}$). In order to explain why the Higgs mass is so small compared to $M_{\mathrm{Pl}}$, supersymmetry postulates that there is a spontaneously broken symmetry between fermions and bosons whereby every boson has a fermionic partner and every fermion has a bosonic partner. Given the difference in sign between the contributions of bosons and fermions to the Higgs mass, this symmetry can stabilize the Higgs mass against quantum corrections~\cite{Maiani:1979cx,Veltman:1980mj,Witten:1981nf,Kaul:1981wp}.

Supersymmetry also provides ingredients necessary for successful axion models. First, the hierarchy between the axion decay constant $f_a$ and $M_{\mathrm{Pl}}$ may be explained by supersymmetry~\cite{Moxhay:1984am,Murayama:1992dj}. Second, supersymmetry naturally predicts the existence of two Higgs doublets, which is exactly the structure required by the DFSZ axion model~\cite{dine_simple_1981,zhitnitsky_possible_1980}.

Incorporating supersymmetry into the PQ extension of the SM introduces the fermionic superpartner of the axion, the axino. In the case of $R$-parity conservation with the axino as the Lightest Supersymmetric Particle (LSP), the axino may appear in the decay chains of heavy supersymmetric particles, opening the possibility of observing an axino signature in collider experiments.

Both the axion and the axino have couplings which are suppressed by $f_{a}$, where $f_{a}$ is inversely proportional to the axion mass, $m_{a}$. Constraints on $f_{a}$ by direct-detection and astrophysical axion searches may be interpreted as constraints on the lifetimes of heavier SUSY particle decays into axinos. Conversely, constraints on the decays of heavier SUSY particles into axinos may, under certain model-dependent assumptions,  be used to constrain $f_{a}$, $m_{a}$, and many of the axion couplings which are probed by direct-detection and astrophysical axion searches. Moreover, in the case of the DFSZ axion where the Higgs sector and therefore the higgsinos are charged under PQ symmetry, the higgsinos contribute to the anomalous axion-photon coupling and may result in a large suppression in this essential probe for many direct detection searches. 

This work focuses on a model which features the decay of next-to-lightest supersymmetric particle (NLSP) higgsino-like states with a mass below 1 TeV into an LSP axino state, since a light higgsino is required for supersymmetry to explain the electroweak hierarchy.\footnote{This is because the quadratic term of the Higgs potential is determined by the sum of the higgsino mass squared and supersymmetry-breaking Higgs mass squared, and if the higgsino is heavy, obtaining the electroweak scale requires fine-tuned cancellation between the two contributions. A light stop and light gluino can also be an essential component of natural SUSY models~\cite{papucciNaturalSUSYEndures2012}, but this has a negligible effect on the processes discussed in this work.}
The search for displaced vertices from higgsinos decaying into axinos is also studied in~\cite{Barenboim:2014kka}.
We estimate the sensitivity to this model of collider searches employing displaced-vertex reconstruction, as well as discuss the implications for and complementarity with direct axion searches and astrophysical constraints. Through a coherent approach that combines both perspectives, we demonstrate that collider experiments are complementary to direct-detection and astrophysical searches by (potentially uniquely) probing parameter space for supersymmetric DFSZ axion models that direct-detection and astrophysical searches cannot yet access.

We begin in section~\ref{sec:axion} with an overview of axion physics. In section~\ref{sec:model} we discuss a supersymmetrized DFSZ axion model, focusing on the dynamics of the neutralinos with the addition of an axino. In section~\ref{sec:collider} we discuss a Monte Carlo simulation model for a signal process which exhibits both displaced vertices and missing transverse momentum. Using a \MadAnalysis{5}~\cite{MadAnalysis5:2012} emulation of the ATLAS experiment at the Large Hadron Collider (LHC)~\cite{ATLAS:MainPaper}, we estimate the sensitivity to this process and projected constraints on this model assuming an integrated luminosity equivalent to the LHC Run 2 dataset. In section~\ref{sec:interpretation} we discuss the interpretation of positive signals and exclusions if we consider the supersymmetric DFSZ axion in the landscape of direct-detection and astrophysical axion searches. We also compare the axino search to searches for light gravitinos which have similar signal processes at colliders.

\section{Overview of axion physics}
\label{sec:axion}
The axion arises as the pseudo-Nambu-Goldstone boson from a new spontaneously broken global $\mathrm{U}(1)_{\mathrm{PQ}}$. The $\mathrm{U}(1)_{\mathrm{PQ}}$ symmetry is anomalous and the axion acquires a coupling to the gluon which modifies the $\theta$-term from eq.~\eqref{eq:theta-term}:
\begin{align}
  \mathcal{L}_{\mathrm{QCD}} \supset \left(\frac{a}{f_{a}} + \theta\right)\frac{g^{2}}{32\pi^{2}}G \tilde{G}.
  \label{eq:theta-mod}
\end{align}
At low energies, the axion interacts with a potential caused by QCD dynamics and obtains a vacuum expectation value (VEV) which cancels the $\theta$-term, thus solving the strong-$CP$ problem dynamically.

The DFSZ axion model provides an explicit UV completion of the effective model just described. It generates the correction to the $\theta$-term by introducing a Higgs doublet to the SM to which we assign PQ charge. This Higgs doublet has Yukawa couplings to the fermions of the SM requiring that these fermions be assigned chiral PQ charge. In particular, through the PQ charged quarks, we obtain a color anomaly of the $U(1)_{\mathrm{PQ}}$ symmetry which gives rise to an effective axion-gluon coupling that cancels the $\theta$-term.

As a result of an electromagnetic anomaly of the PQ symmetry as well as mixing between the axion and neutral mesons at low energies, the axion obtains an effective coupling to photons~\cite{bae2017prospects}:
\begin{align}
\label{eq:axion-photon}
\mathcal{L}_{\mathrm{axion}} & \supset \frac{1}{4} g_{a\gamma\gamma}aF\tilde{F}, \\
\label{eq:photon-coupling}
g_{a\gamma\gamma} &= \frac{\alpha}{2\pi}\frac{1}{f_a}\left(\frac{E}{N} - \frac{2}{3}\frac{4 + z}{1 + z}\right),
\end{align}
where $z = m_u/m_d$~\cite{bae2017prospects} and $E/N$ is the ratio of the electromagnetic anomaly to the color anomaly. 
Eq.~\eqref{eq:axion-photon} causes the conversion of axions to photons in the presence of a strong magnetic field. By measuring the rate of conversion of axions into photons (or the lack thereof) in a strong magnetic field, constraints can be placed on $g_{a\gamma\gamma}$ as well as $f_{a}$. Moreover, in order to solve the strong-CP problem, the QCD axion has a mass which is related to $f_{a}$~\cite{bae2017prospects} as:
\begin{align}
\label{eq:axion-mass}
m_{a} &= \frac{z^{1/2}}{1 + z} \frac{f_{\pi}m_{\pi}}{f_{a}} \simeq 6 \ \mu\mathrm{eV} \frac{10^{12} \ \mathrm{GeV}}{f_{a}}.
\end{align}
The axion-photon coupling of the form eq.~\eqref{eq:axion-photon} exists not only for the QCD axion, but for a broad class of axion-like particles (ALPs). Due to the generic nature of eq.~\eqref{eq:axion-photon}, direct detection constraints on $g_{a\gamma\gamma}$ -- and therefore also $f_{a}$ -- are relatively model-independent and can simultaneously constrain both QCD axions and ALPs with a non-zero photon coupling. Other axion couplings may be probed such as the axion-electron or axion-nucleon couplings. However, those couplings are often more model-dependent as they can have tree level contributions from the UV theory as well as more generic contributions which enter at lower energies. In section~\ref{sec:interpretation}, we will assess constraints from astrophysical observations on these other couplings which arise from the specific supersymmetrized DFSZ model considered in this work.

In the case of the supersymmetric DFSZ axion, the addition of the fermionic higgsinos changes the electric and color anomalies, resulting in $E / N = 2$ in the anomalous contribution to the axion-photon coupling. This leads to an axion-photon coupling which lies a factor of 20 below the non-supersymmetric DFSZ axion coupling, or, at the cost of fine-tuning, which completely vanishes~\cite{bae2017prospects}. (See~\cite{Badziak:2023fsc} for an analysis including higher-order chiral perturbations, which find that the possibility of cancellation persists.) In the case that such a suppression of the axion-photon coupling is realized, the collider-based approaches discussed here -- as well as astrophysical observations -- may be the best way to probe the supersymmetric DFSZ axion until direct-detection searches are able to reach the requisite sensitivities.

\section{Axion \& axino model}
\label{sec:model}
Supersymmetry alone does not solve the hierarchy problem completely.
The supersymmetric Higgs mass parameter $\mu$ in the minimal supersymmetric SM (MSSM) must be taken to be around the electroweak scale and be much below other fundamental scales such as the Planck scale. This may be naturally achieved if $\mu$ is forbidden by a symmetry that is spontaneously broken.

The DFSZ axion gives the Higgs doublets PQ charges (as well as the particles which obtain their masses from the Higgs). The $\mu$ term is forbidden by the PQ symmetry and instead the following PQ-invariant superpotential is introduced, 
\begin{align}
\label{eq:DFSZ-PQMSSM}
W_{\mathrm{DFSZ-PQMSSM}} \supset \lambda \frac{S^{n}}{M_{\mathrm{Pl}}^{n-1}} H_{u}H_{d},
\end{align}
where $S$ is a gauge-singlet PQ-charged field.
The $\mu$ term is dynamically generated when the PQ symmetry is broken, with  $\mu = \lambda\left\langle S \right\rangle^{n}/{M_{\mathrm{Pl}}^{n-1}} \sim f_{a}^{n}/{M_{\mathrm{Pl}}}^{n-1}$. For $f_a \ll M_{\mathrm{Pl}}$, we obtain a $\mu$ parameter which is around the proper electroweak scale, as in the Kim-Nilles model~\cite{KIM1984150}. We will refer to the MSSM with the gauge singlet field $S$, the global PQ symmetry, and the interaction in eq.~\eqref{eq:DFSZ-PQMSSM}, as the DFSZ-PQMSSM.

After PQ symmetry breaking, expanding about the potential of the gauge singlet field, the Kim-Nilles term gives:
\begin{align}
  W_{\mathrm{DFSZ-PQMSSM}} &\supset \mu H_{u}H_{d} + y_{a}A H_{u}H_{d} \; ,
  \label{eq:superpotential} \\
\text{with} \quad y_{a} &= \frac{\sqrt{2}n}{N_{\mathrm{DW}}}\frac{\mu}{f_{a}} \; ,
\end{align}
Here $N_{\mathrm{DW}}$ is the axion domain wall number. We write the superpotential in terms of $\mu$ when possible rather than the more fundamental parameters of our model because $\mu$ has more immediate phenomenological relevance. Without extra coupling of $S$ to colored particles, $N_{\mathrm{DW}} = 3n$ and the Yukawa coupling is solely determined by $\mu$ and $f_{a}$.

In the above, $A = \left\{\frac{1}{\sqrt{2}}(s + ia), \tilde{a}\right\}$ is the axion supermultiplet containing the scalar saxion, $s$, the pseudoscalar axion, $a$, and the axino, $\tilde{a}$. The saxion is introduced in supersymmetric theories because chiral superfields require an extra degree of freedom which the real axion field alone does not have. The axion corresponds to oscillations in the approximately flat phase degree of freedom in the wine-bottle PQ breaking potential whereas the saxion corresponds to oscillations in the radial degree of freedom.

\begin{figure}[h]
  \centering
  \includegraphics[width=0.8\textwidth]{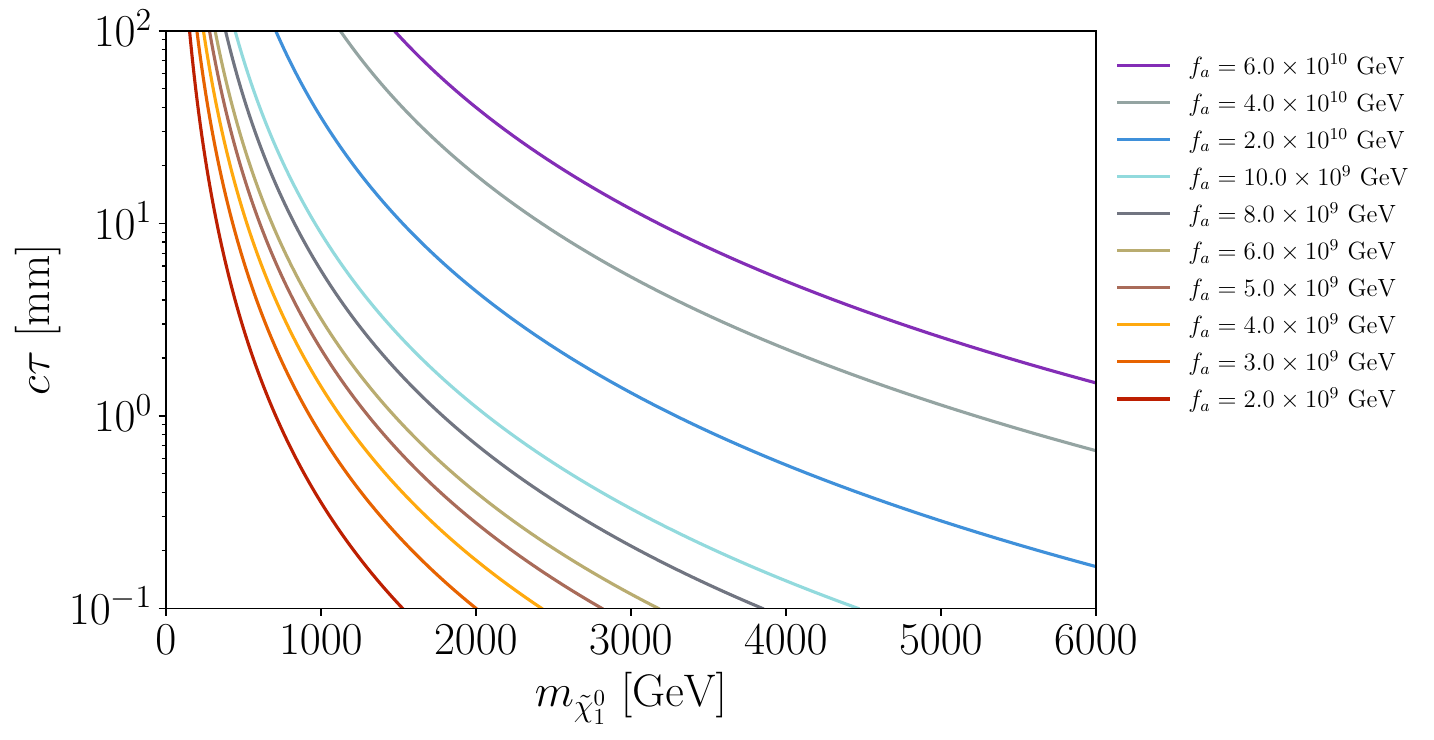}
  \caption{\label{fig:rainbow_lifetime} Higgsino NLSP lifetime, $c\tau$, dependence on the higgsino mass, $m_{\tilde{\chi}_{1}^{0}}$, for various axion decay constants $f_{a}$.}
\end{figure}

The Kim-Nilles term thus induces Yukawa couplings between the Higgs sector and the axion/saxion/axino after PQ symmetry is broken. Moreover, after electroweak symmetry breaking, the axino mixes with the neutralinos of the MSSM.
In the case where $R$-parity is conserved, all heavier supersymmetric particles must decay to lighter supersymmetric particles meaning that, if the axino is relatively light, it must appear in the decays of heavier supersymmetric particles. Although axion couplings are too feeble to be seen in collider experiments, the axino may be observable because $R$-parity conservation would require that the axino appear in decays of heavier supersymmetric particles. The $\propto 1/f_{a}$ suppression will manifest for the axino as a lifetime of the decays of heavier particles into an axino~\cite{Redino:2015mye}. The lifetime of the higgsino-like NLSP is given approximately by the following expression:
\begin{align}
  c\tau \approx \mm{8.9} \left(\frac{f_{a}}{10^{10} \, \gevonly}\right)^{2} \left(\frac{\tev{1}}{m_{\tilde{\chi}_{1}^{0}}}\right)^{3}.
  \label{eq:lifetime_approx}
\end{align}
Figure~\ref{fig:rainbow_lifetime} and eq.~\eqref{eq:lifetime_approx} show the scaling of the NLSP lifetime, $c\tau$, with the square of $f_{a}$ and with the inverse of the cube of $m_{\tilde{\chi}_{1}^{0}}$, resulting in a significant fraction of parameter space that yields potentially detectable signatures at the LHC.

After electroweak symmetry breaking, the axino shares the same charges of the unbroken symmetries as the neutralinos of the MSSM which allows for mixing between the neutralinos and the axino. This mixing is implemented as a $5 \times 5$ unitary mixing matrix where an additional row and column are added to the neutralino mixing matrix of the MSSM to account for the additional axino state~\cite{choi2001analysis, bae2012cosmology}. The scalars and pseudoscalars of the model also mix such that the standard model Higgs is primarily a mixture of the neutral scalar components of the Higgs doublets like in the MSSM, but there is also a negligible contribution from the saxion. In the case where the two heaviest neutralinos are mostly wino and bino, and they are much heavier than the lighter neutralino states, the mixing between the axino and mostly higgsino states and the heavier neutralinos is negligible. In the mass basis, we have heavier neutralino states $\tilde{\chi}_{3}^{0}$ and $\tilde{\chi}_{4}^{0}$ which consist mostly of electroweak gauginos, $\tilde{\chi}_{1}^{0}$ and $\tilde{\chi}_{2}^{0}$ which consist mostly of the neutral higgsinos with $\tilde{\chi}_{1}^{0}$ taken to be the NLSP, and an approximately pure axino state which we take to be the LSP.
\begin{align}
\begin{pmatrix}
\tilde{\chi}_{4}^{0}\\
\tilde{\chi}_{3}^{0}\\
\tilde{\chi}_{2}^{0}\\
\tilde{\chi}_{1}^{0}\\
\tilde{a}
\end{pmatrix} \approx N\begin{pmatrix}
\tilde{B}\\
\tilde{W}^{3}\\
\tilde{H}_{d}^{0}\\
\tilde{H}_{u}^{0}\\
\tilde{a}
\end{pmatrix}.\end{align}
The derivation of the mixing matrix is described in more detail in the appendix~\ref{sec:appendix}. The mixing matrix should diagonalize the neutralino mass matrix:
\begin{align}\mathcal{M}_{\mathrm{diag}} = N^{*}\mathcal{M} N^{\dagger}.\end{align}
In the limit where the electroweak gaugino masses are much larger than $m_{Z}$, we can perturbatively diagonalize the mass matrix to get an approximate mixing matrix~\cite{beylin2008diagonalization, bae2012cosmology}. The mixing matrix gives us axino mixings which are suppressed by $y_{a} \sim 1/f_{a}$. We also have gaugino mixings which are suppressed by $\frac{m_{Z}}{M_{1}}$ and $\frac{m_{Z}}{M_{2}}$ which are small if the mostly gaugino neutralino states are taken to be heavy.
The masses of $\tilde{\chi}_{1}^{0}$ and $\tilde{\chi}_{2}^{0}$ are~\cite{Martin:1997ns,beylin2008diagonalization, bae2012cosmology}
\begin{align}
m_{\tilde{\chi}_{1}^{0}} \approx \mu + \mathcal{O}\left(\frac{m_{Z}^2}{M_{1}}\right) + \mathcal{O}\left(\frac{m_{Z}^2}{M_{2}}\right) + \mathcal{O}\left(y_{a}\right)
  \label{eq:splitting1}
\end{align}
\begin{align}
m_{\tilde{\chi}_{2}^{0}} \approx -\mu + \mathcal{O}\left(\frac{m_{Z}^2}{M_{1}}\right) + \mathcal{O}\left(\frac{m_{Z}^2}{M_{2}}\right) + \mathcal{O}\left(y_{a}\right).
  \label{eq:splitting2}
\end{align}
So, for large values of $M_{1}$, $M_{2}$ and $f_{a}$, the difference in mass between the two mostly higgsino states becomes small.

The lightest chargino state is approximately degenerate in mass with $\tilde{\chi}_{1}^{0}$ and $\tilde{\chi}_{2}^{0}$. The chargino mass matrix has the form eq.~\eqref{eq:chargino-matrix}. As is shown in the appendix, this can be diagonalized to yield a lighter chargino mass of:
\begin{align}
m_{\tilde{\chi}_{1}^{\pm}} \simeq & \mu - \mathcal{O}\left(\frac{m_{W}^{2}}{M_{2}}\right).
\end{align}
So, in the limit where $M_{2}$ is taken to be large, the mass of $\tilde{\chi}_{1}^{\pm}$ is roughly degenerate with the masses of $\tilde{\chi}_{1}^{0}$ and $\tilde{\chi}_{2}^{0}$.

\section{Collider signatures}
\label{sec:collider}
\begin{figure}[h]
  \centering
    \includegraphics[width=0.4\textwidth]{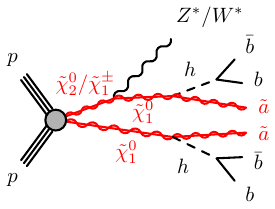}
    \includegraphics[width=0.4\textwidth]{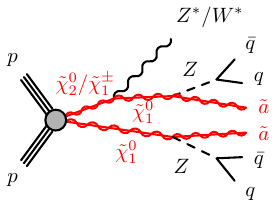}
    \caption{\label{fig:N1N2axino} Feynman diagrams showing chargino-neutralino or neutralino-neutralino pair production from a proton-proton collision. In both cases, due to the small mass splitting, the heavier $\tilde{\chi}_{2}^{0}$ or $\tilde{\chi}_{1}^{\pm}$ state first decays to the lightest neutralino $\tilde{\chi}_{1}^{0}$ by the emission of an extremely off-shell $Z$ or $W$ boson, respectively. Both $\tilde{\chi}_{1}^{0}$ then decay to the axino, the lightest supersymetric particle, by the emission of a Higgs boson (left) or $Z$ boson (right). Supersymmetric particles are shown in red.}
\end{figure}

This DFSZ-PQMSSM model gives rise to several signatures that are detectable at an experiment like ATLAS or CMS~\cite{CMS:2008xjf} at the LHC; figure~\ref{fig:N1N2axino} shows two representative Feynman
diagrams. As shown in eq.~\eqref{eq:splitting1} and eq.~\eqref{eq:splitting2}, the difference in mass between the heavier and
lighter mostly Higgsino states is small in the limit of large $M_{1}$ and $M_{2}$. This holds for both $\tilde{\chi}_{1}^{0}$
and $\tilde{\chi}_{2}^{0}$, as well as $\tilde{\chi}_{1}^{0}$ and $\tilde{\chi}_{1}^{\pm}$, which will be mostly the
charged Higgsino state. These heavier Higgsino states will primarily decay down to $\tilde{\chi}_{1}^{0}$ by the
emission of $Z$ or $W$ boson, respectively.  
This tree-level decay is suppressed by the small mass splitting. To be concrete, in this work, we take $M_{1}, M_{2} \sim 1 \ \mathrm{TeV}$, which leads to a small
mass splitting but a prompt decay of the heavier $\tilde{\chi}_{2}^{0}$ or $\tilde{\chi}_{1}^{\pm}$ into an off-shell $W$ or $Z$ boson and the lighter $\tilde{\chi}_{1}^{0}$ higgsino state. The extra $Z$ or $W$ boson
from this decay will be very offshell and is likely much too soft to be reconstructed.%
\footnote{
If $M_1$ and $M_2$ are much above the TeV scale, $\tilde{\chi}_{2}^{0}$ directly decays into $Z/h$ and the axino with the same rate as $\tilde{\chi}_{1}^{0}$ , and the signal remains essentially the same. $\tilde{\chi}_{1}^{\pm}$ is heavier than $\tilde{\chi}_{1}^{0}$ by about 300 MeV via electroweak corrections and can decay into $\tilde{\chi}_{1}^{0}$ with a decay length of 10 mm~\cite{Ibe:2023dcu}. $\tilde{\chi}_{1}^{\pm}$ may then dominantly decay into $W^{\pm}$ and the axino, and jets come from $W$ rather than $h/Z$. Even in this case, the signal is similar to $h/Z + \tilde{a}$ and we expect a similar sensitivity.}

What would be observable is the subsequent decay of both neutralinos to axinos via the emission of either a Higgs
or $Z$ boson. This decay is suppressed by $1/f_a$, and can easily cause the neutralino to become long-lived for
a range of values of $f_a$ (see eq.~\eqref{eq:lifetime_approx}). Depending on the lifetime, different types of long-lived particle (LLP) searches
may have sensitivity to this model. If the lifetime is reasonably short ($\tau \sim$ nanoseconds), an LLP
will decay in the charged-particle trackers of the LHC experiments. If the LLP decays result in charged-particles in the final state, a reconstructible decay vertex commonly known as a ``displaced vertex'' may provide a distinctive signature. Searches for such displaced vertices are often nearly background free~\cite{ATLAS:2017tny}, giving them significant potential to discover even very rare processes. In this work, we therefore focus on the case where the Higgs or $Z$ bosons decay hadronically (to $b\bar{b}$
specifically in the case of the Higgs, as it has by far the highest cross section); we also limit ourselves to the
range of $f_a$ values that will lead to a displaced vertex. In the case where a chargino appears in the signal process, the chargino may leave a reconstructible track before decaying to soft particles and the long-lived NLSP creating a ``disappearing track.'' Accounting for such a feature may provide higher sensitivity for searches at new colliders as we will mention briefly in section~\ref{sec:interpretation}, but we will not discuss disappearing tracks in detail beyond that.

The axino is both neutral and stable as the LSP, and so escapes the detector, resulting in large missing transverse momentum, $E_{\mathrm{T}}^{\mathrm{miss}}$, defined as the negative vector sum of the transverse momenta of all reconstructed physics objects. If one of the neutralinos is sufficiently long-lived and decays outside the fiducial volume of the charged-particle tracker, it will also contribute to the overall $E_{\mathrm{T}}^{\mathrm{miss}}$. The detector signature for chargino-chargino, chargino-neutralino, and neutralino-neutralino pair production is therefore the same, one or more displaced vertices along with large $E_{\mathrm{T}}^{\mathrm{miss}}$, and the decay products of the $h$/$Z$.

In order to estimate the sensitivity of a search for this signature at a collider experiment, the model described above was implemented in the \SARAH framework~\cite{SARAH:2012,SARAH4:2014} and exported to a Universal Feynrules Output~\cite{UFO:2011} (UFO) format file. Using the resulting UFO, truth-level Monte Carlo events for processes like the one shown in figure~\ref{fig:N1N2axino} were generated for different values of $f_{a}$, $m_{\tilde{a}}$, and $m_{\tilde{\chi}_{1}^{0}}$, using \Madgraph 2.9.16~\cite{Madgrah:2014} with NNPDF2.3 leading-order parton distribution function set~\cite{PartonDistributions:2012}; particle decays were handled with \Madspin~\cite{MadSpin:2012} and \Pythia~\cite{Pythia8.3:2022} with the A14 tune~\cite{AtlasA14:2014}. To simulate detector effects and implement selection criteria, a background-free analysis was implemented using the \MadAnalysis{5} framework~\cite{MadAnalysis5:2012} which can apply cuts and detector effects to our truth-level Monte Carlo events~\cite{araz2021simplified}. Within the \MadAnalysis{5} framework, jets were reconstructed using \Fastjet~\cite{FastJet:2012}. In this case, a fast detector simulation within the \MadAnalysis{5} framework based on the ATLAS detector was chosen. An existing \MadAnalysis{5} implementation of a different ATLAS displaced vertex analysis, looking for oppositely-charged leptons as final states~\cite{DVN/31JVGJ_2021}, was used as a starting point for this study.

ATLAS has also previously searched for displaced vertex signatures with $E_{\mathrm{T}}^{\mathrm{miss}}$ in the final state~\cite{ATLAS:2017tny}. Following this analysis, we apply a preselection on potential signal processes requiring $m_{\mathrm{DV}} > \gev{10}$ and $N_{\mathrm{track}} \geq 5$~\cite{ATLAS:2017tny}. Additionally, we require that the vertex be displaced by at least \mm{4} from all primary vertices, must be in the fiducial volume of the tracker ($R < \mm{300}$ and $|z| < \mm{300}$), and is not in regions of the detector which are disabled or which are too rich in material~\cite{ATLAS:2017tny}. Vertices which pass the preselection criteria then have reconstruction efficiencies which are taken from the previous ATLAS search~\cite{ATLAS:2017tny}, and which depend on the vertex mass, radius, and the number of tracks used to reconstruct the vertex. The $E_{\mathrm{T}}^{\mathrm{miss}}$ component of the signal is implemented as a cut and we only consider signal processes with $E_{\mathrm{T}}^{\mathrm{miss}} > \gev{150}$.

\begin{figure}[h]
  \centering
  \subfloat[$m_{\tilde{\chi}_{1}^{0}} = \gev{600}$, $m_{\tilde{a}} = \gev{300}$, varying $f_{a}$]{\includegraphics[width=0.45\textwidth]{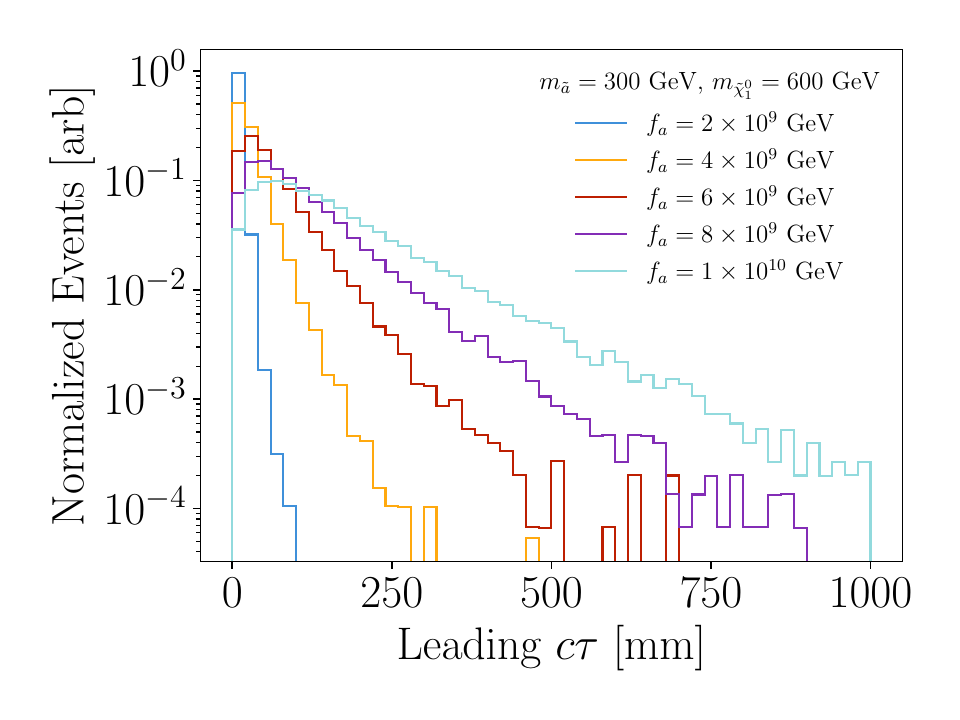}}
  \subfloat[$f_{a} = 5 \times 10^{9} \gevonly$, varying $m_{\tilde{\chi}_{1}^{0}}$]{\includegraphics[width=0.45\textwidth]{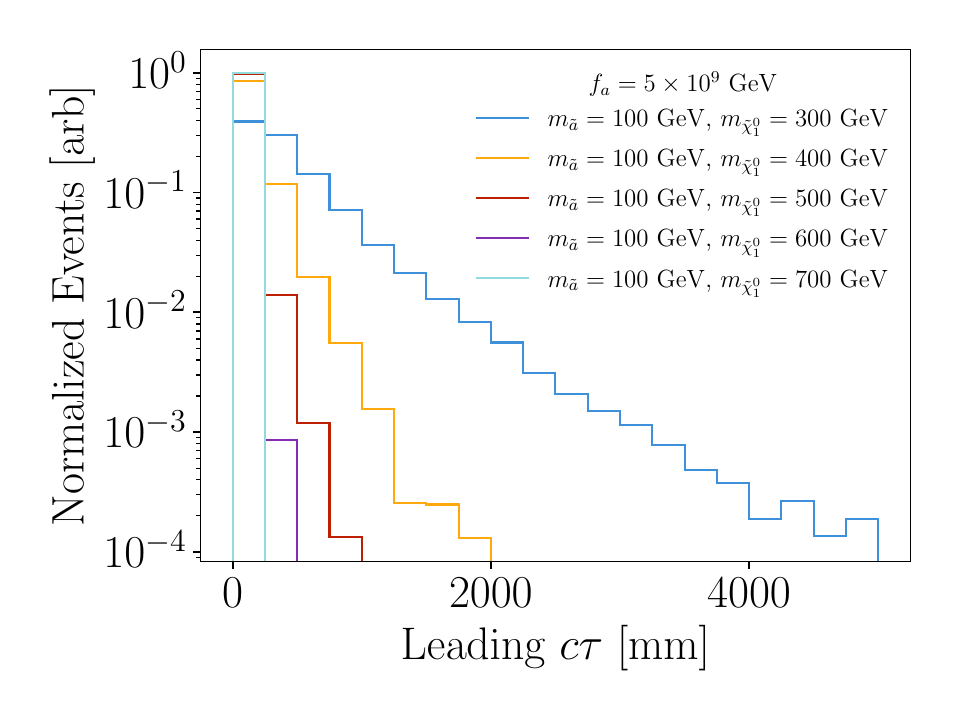}}
  \caption{\label{fig:lifetime} Distributions of the decay length ($c\tau$) from the longest lived NLSP state in an event for different model parameters. These distributions are taken before any preselection criteria are applied.}
\end{figure}

\begin{figure}[h]
  \centering
  \subfloat[$m_{\tilde{\chi}_{1}^{0}} = \gev{600}$, $m_{\tilde{a}} = \gev{300}$, varying $f_{a}$]{\includegraphics[width=0.45\textwidth]{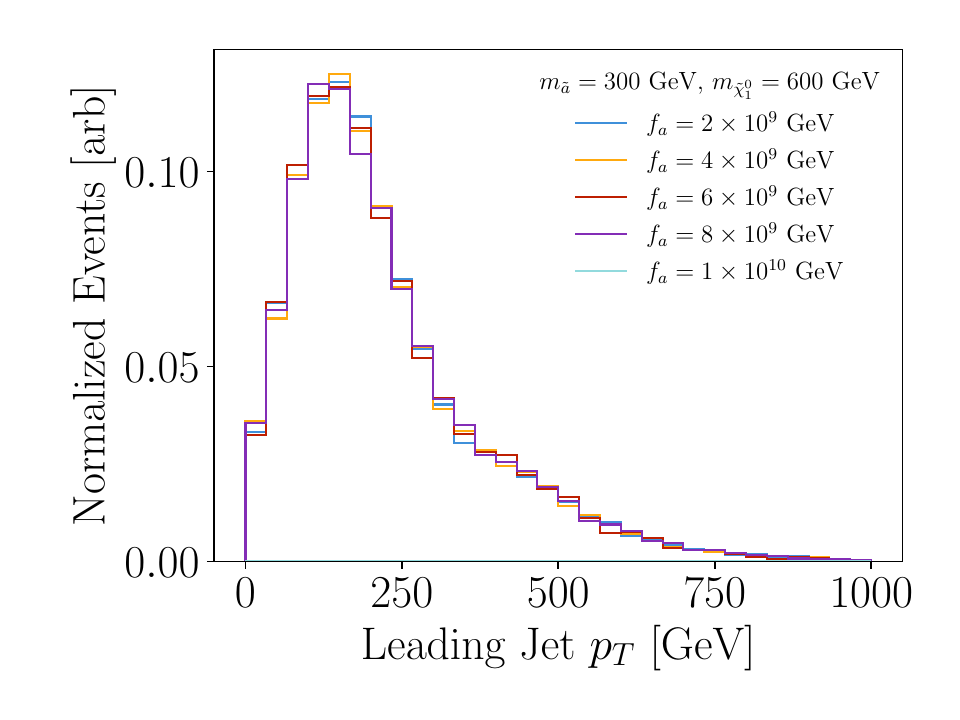}}
  \subfloat[$f_{a} = 5 \times 10^{9} \gevonly$, varying $m_{\tilde{a}}$]{\includegraphics[width=0.45\textwidth]{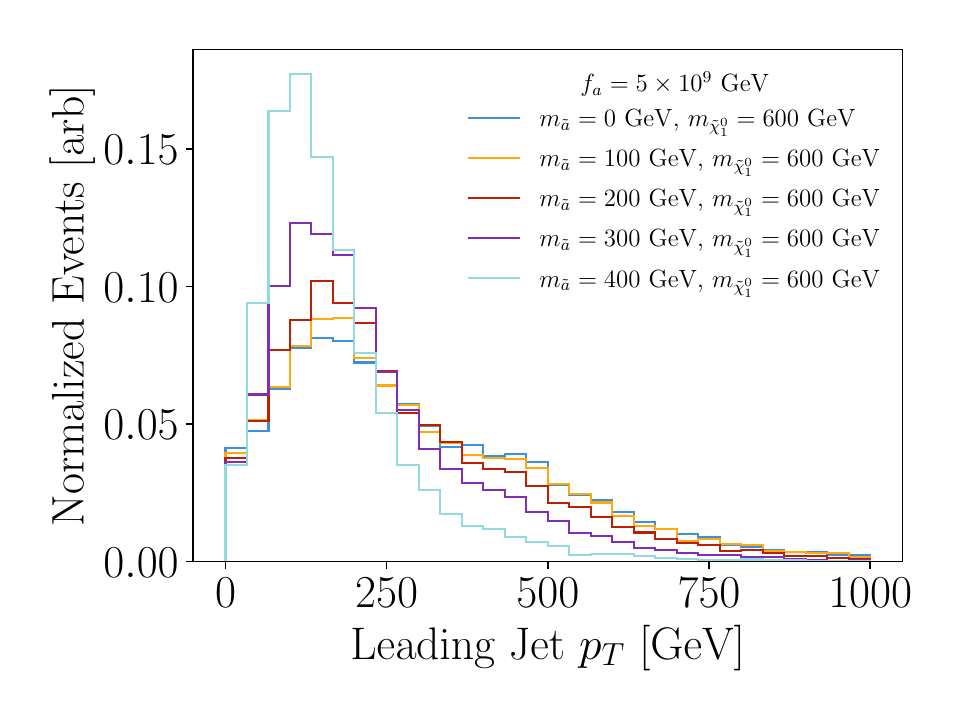}}
  \caption{\label{fig:jlpT} Distributions of leading jet transverse momentum for different model parameters. These distributions are taken before any preselection criteria are applied.}
\end{figure}

\begin{figure}[h]
  \centering
  \subfloat[$m_{\tilde{\chi}_{1}^{0}} = \gev{600}$, $m_{\tilde{a}} = \gev{300}$, varying $f_{a}$]{\includegraphics[width=0.45\textwidth]{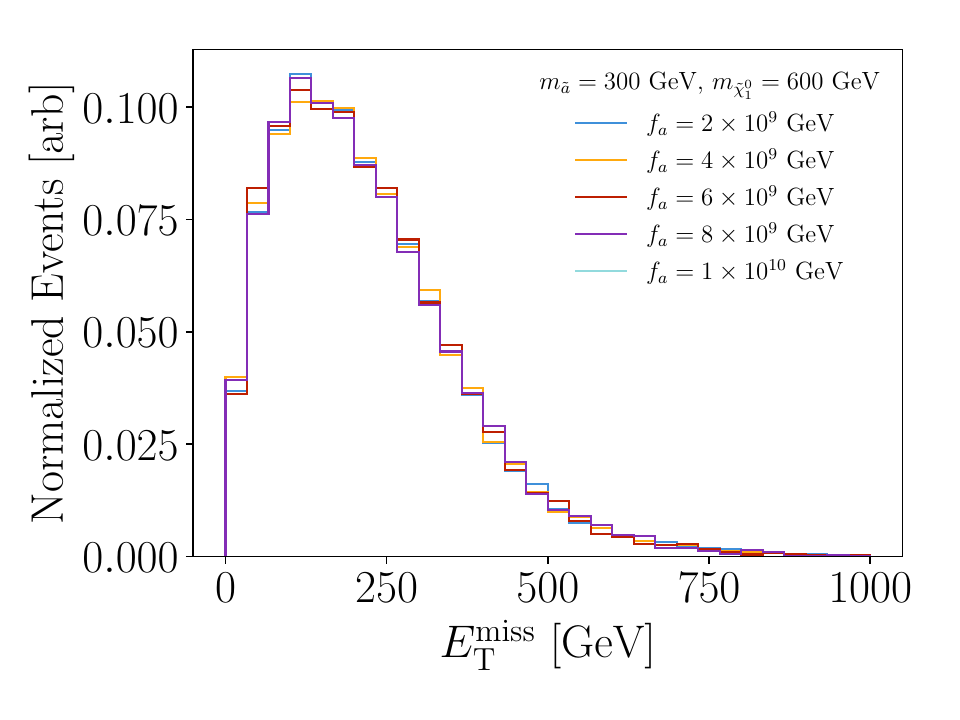}}
  \subfloat[$f_{a} = 5 \times 10^{9} \gevonly$, varying $m_{\tilde{a}}$]{\includegraphics[width=0.45\textwidth]{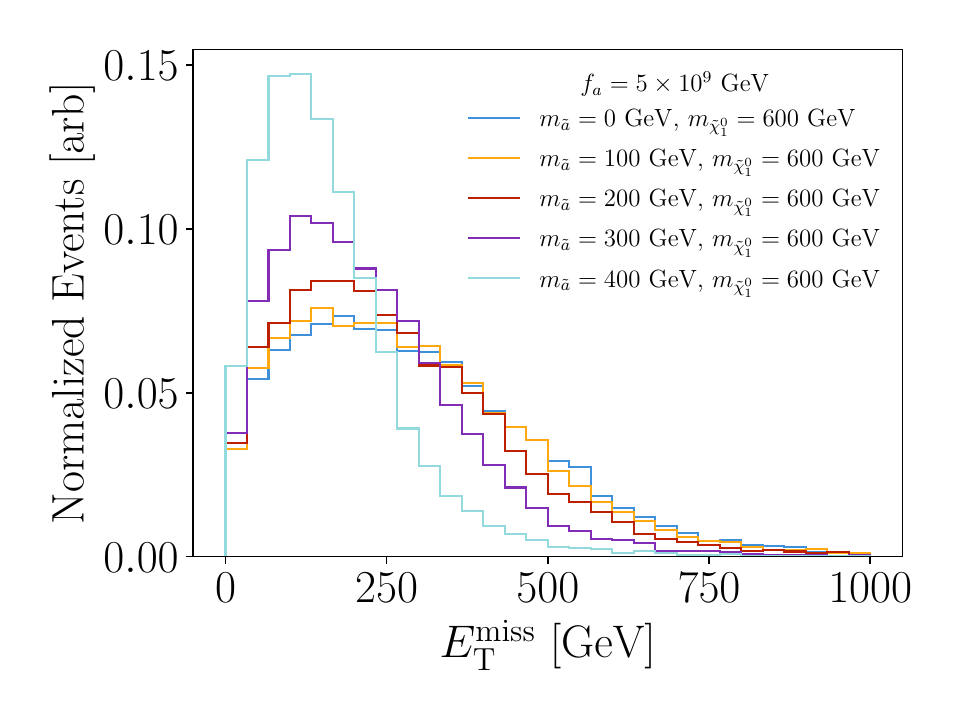}}
  \caption{\label{fig:MET} Distributions of $E_{\mathrm{T}}^{\mathrm{miss}}$ for different model parameters. These distributions are taken before any preselection criteria are applied.}
\end{figure}

Figure~\ref{fig:lifetime} shows the distribution of the decay length for the longest lived NLSP in each event. For increasing $f_{a}$ values the leading NLSP lifetime also increases which is consistent with the $1/f_{a}$ suppression in the coupling in the superpotential eq.~\eqref{eq:superpotential} for this model. We also observe that heavier NLSP corresponds to shorter NLSP lifetime due to the increased range of final momentum states accessible to heavier decays.

Figure~\ref{fig:jlpT} shows the distributions of the leading jet transverse momentum for the generated Monte Carlo events. Figure~\ref{fig:MET} shows the distributions of $E_{\mathrm{T}}^{\mathrm{miss}}$ which were reconstructed in \MadAnalysis{5}. The kinematic distributions are unaffected by scaling $f_{a}$ because this primarily changes the lifetime of the particle. Scaling $f_{a}$ can make the lifetime larger or smaller which may affect the reconstruction efficiency of the displaced vertices, but does not significantly change the distributions of leading jet $p_{\mathrm{T}}$ or $E_{\mathrm{T}}^{\mathrm{miss}}$. Figure~\ref{fig:jlpT} and figure~\ref{fig:MET} also show that lighter axinos correspond to higher leading jet $p_{\mathrm{T}}$ and higher $E_{\mathrm{T}}^{\mathrm{miss}}$. This is because lighter axino states will have comparatively more 3-momentum, leading to higher 3-momentum in visible final state particles and higher momentum transfer to the axino state.

\begin{figure}[h]
  \centering
    \captionsetup[subfigure]{labelformat=empty}
    \subfloat
    {
        \includegraphics[width=0.45\textwidth]{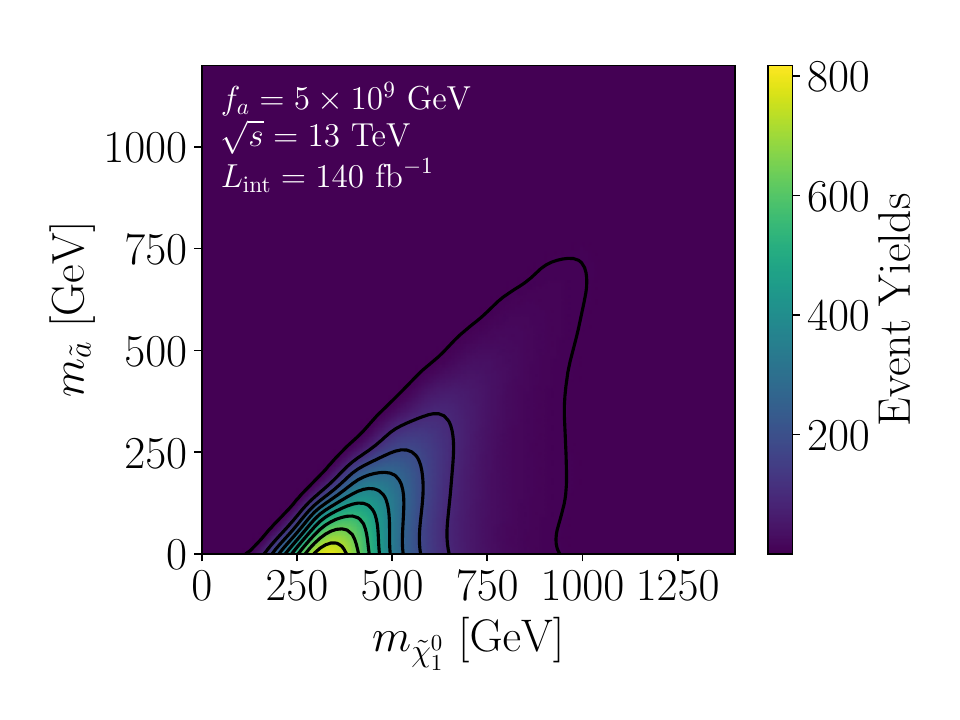}
    }
    \subfloat
    {
        \includegraphics[width=0.45\textwidth]{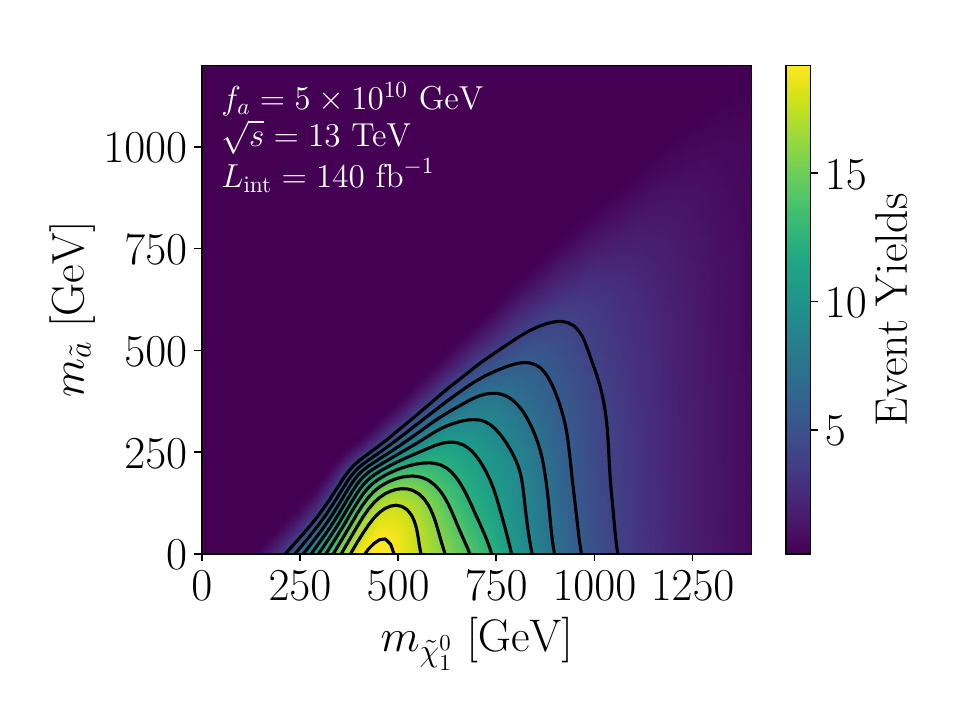}
    }
    \caption{
        \label{fig:yields} Event yield contours in the $m_{\tilde{\chi}_1^0}-m_{\tilde{a}}$ plane for two different values of $f_{a}$. The contour lines and color map are derived from an interpolation of individual points in parameter space.
    }
\end{figure}

\begin{figure}[h]
  \centering
    \captionsetup[subfigure]{labelformat=empty}
    \subfloat
    {
        \includegraphics[width=0.45\textwidth]{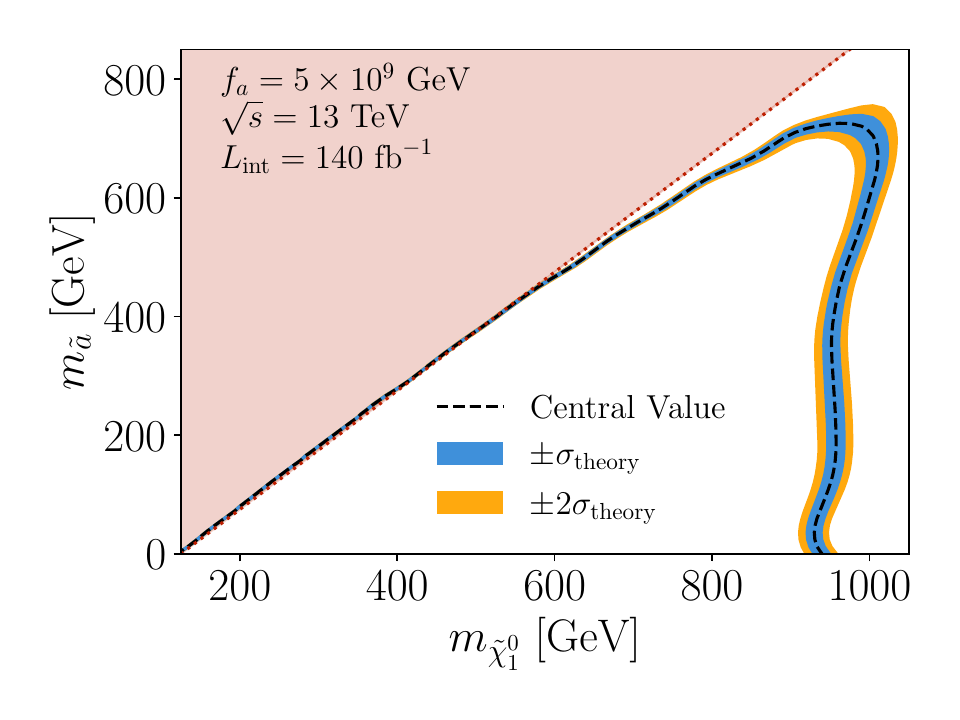}
    }
    \subfloat
    {
        \includegraphics[width=0.45\textwidth]{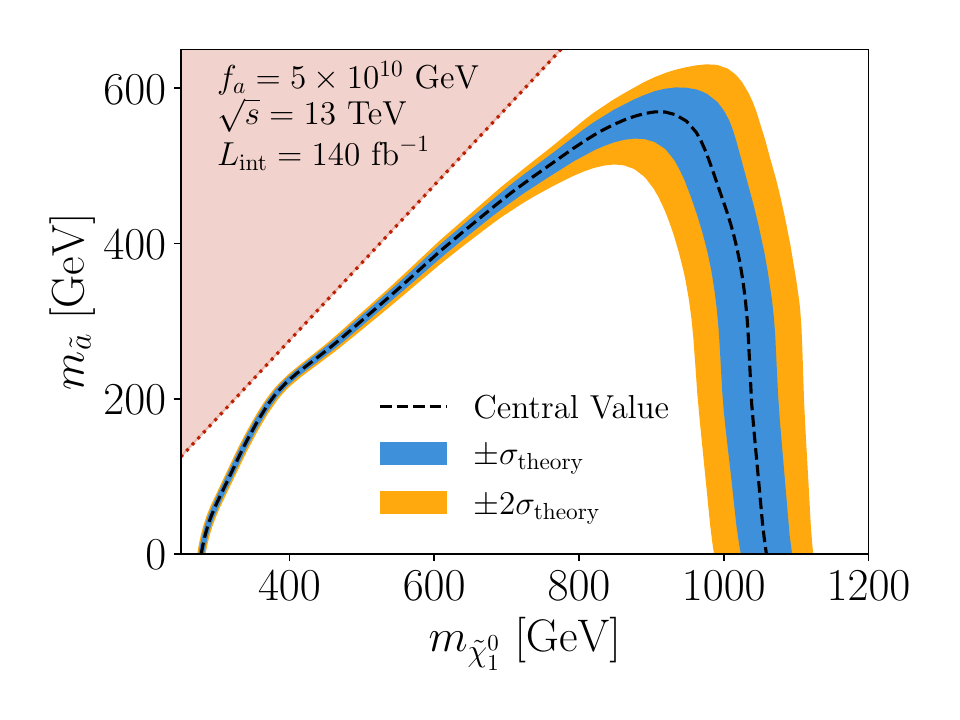}
    } 
    \caption{
        \label{fig:exclusion-contour} Determination of the allowed parameter space in the $m_{\tilde{\chi}_1^0}-m_{\tilde{a}}$ plane at $95\%$ confidence level. The theory uncertainties shown in these plots are derived from the uncertainty in the NLO higgsino pair production cross section~\cite{Fuks:2012qx, Fuks:2013vua}. The red shaded region indicates kinematically forbidden parameter space.
    }
\end{figure}

The heavier NLSP states can decay into either a higgs and an axino or a $Z$ boson and an axino. For the former case, the signal yields obtained after applying kinematic selection criteria and detector reconstruction filters are scaled using the branching ratios $\mathrm{BR}^{2}(\tilde{\chi}_{1}^{0} \to \tilde{a}h)$ and $\mathrm{BR}^{2}(h \to b\overline{b})$~\cite{CERN_4}, the appropriate NLO SUSY cross section\footnote{A more recent calculation of these cross sections with higher precision can be found in~\cite{fiaschi2020higgsino}. These updated cross sections have negligible effect on the results shown in this work.} for pair production of a NLSP with a given higgsino mass~\cite{Fuks:2012qx, Fuks:2013vua}, and an integrated luminosity which is taken to be $140 \ \mathrm{fb}^{-1}$. Similar scaling is performed on events in which $Z$ bosons are produced in the NLSP decay, but, since the acceptance was determined to be roughly the same, the $\tilde{\chi}_{1}^{0} \to \tilde{a}h$ events were used to represent both types of signal events and estimate the overall sensitivity to model parameters. The interpolated contours giving the event yields for different model parameters are shown in figure~\ref{fig:yields}. The yields are higher for lighter axinos which allow for larger $E_{\mathrm{T}}^{\mathrm{miss}}$ and leading jet $p_{\mathrm{T}}$ as discussed above and shown in figure~\ref{fig:jlpT} and figure~\ref{fig:MET}. The yields also vary based on the mass of the NLSP as that affects the NLSP lifetime as described by eq.~\eqref{eq:lifetime_approx}. If the NLSP is too light it may be so long-lived that it decays outside of the detector and if the NLSP is too heavy it may decay so quickly that it does not yield a displaced vertex. From the expected event yields, we determined the exclusion sensitivities at $95\%$ confidence level shown in figure~\ref{fig:exclusion-contour}. The theory uncertainties in figure~\ref{fig:exclusion-contour} are calculated from the uncertainty in the higgsino pair production cross section~\cite{Fuks:2012qx, Fuks:2013vua}. Due to the requirement that the NLSP decay produce an on-shell higgs, there is a kinematically forbidden region for higgsino masses $m_{\tilde{\chi}_{1}^{0}} < m_{\tilde{a}} + m_{h}$.

\begin{figure}[h]
  \centering
    \captionsetup[subfigure]{labelformat=empty}
    \subfloat[
    ]
    {
        \includegraphics[width=0.45\textwidth]{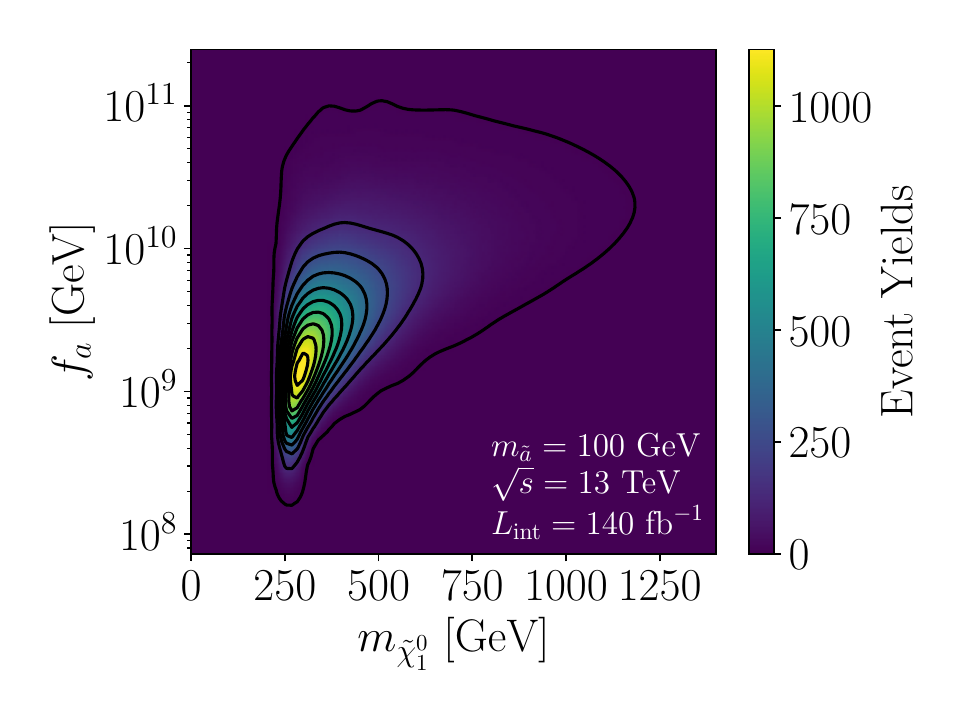}
    }
    \subfloat[
    ]
    {
        \includegraphics[width=0.45\textwidth]{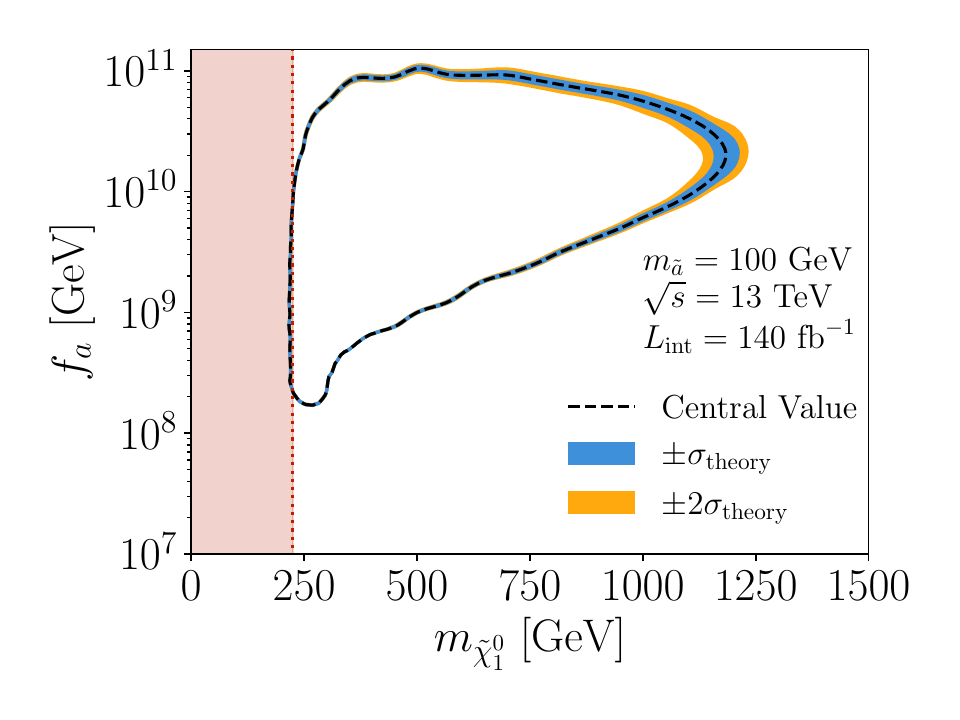}
    } 
    \caption{
        \label{fig:stats-maxino-slice} An interpolated contour plot showing the projected event yields in the $m_{\tilde{\chi}_1^0}-f_a$ plane (left) and an exclusion contour showing the allowed parameter space in the $m_{\tilde{\chi}_{1}^{0}}-f_{a}$ plane at $95\%$ confidence level (right). The theory uncertainties shown in these plots are derived from the uncertainty in the higgsino production cross section~\cite{Fuks:2012qx, Fuks:2013vua}. The red shaded region indicates kinematically forbidden parameter space.
    }
\end{figure}

\begin{figure}[h]
  \centering
  \includegraphics[width=0.8\textwidth]{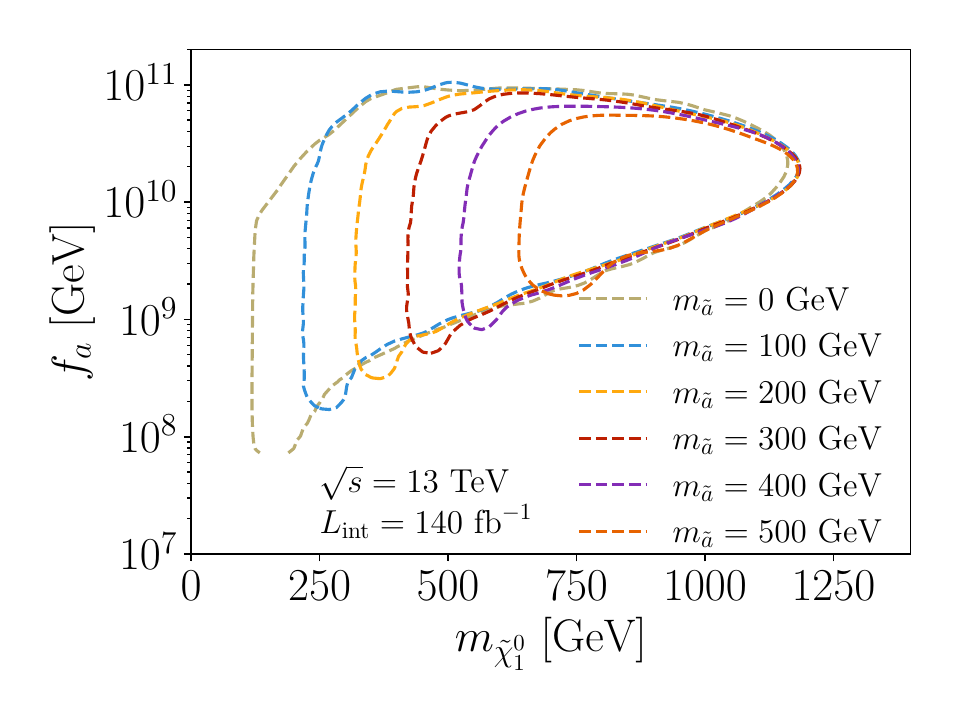}
  \caption{\label{fig:maxino-slice-exclusions-combined} Projected exclusions of the $m_{\tilde{\chi}_{1}^{0}}-f_{a}$ parameter space at $95\%$ confidence level for different values of $m_{\tilde{a}}$. For simplicity, theory uncertainties are omitted and only central values are shown.}
\end{figure}

Simulated events were also generated on a grid in the $m_{\tilde{\chi}_{1}^{0}}-f_{a}$ plane for fixed values of $m_{\tilde{a}}$. An interpolation of the event yields obtained from \MadAnalysis{5} and the corresponding projected exclusion at $95\%$ confidence level are shown in figure~\ref{fig:stats-maxino-slice}. As in figure~\ref{fig:exclusion-contour}, the theory uncertainties shown in figure~\ref{fig:stats-maxino-slice} are derived from uncertainties in the NLO cross section for higgsino pair production~\cite{Fuks:2012qx, Fuks:2013vua}. As in figure~\ref{fig:exclusion-contour}, there is a kinematically forbidden region due to the requirement of an on-shell higgs: $m_{\tilde{\chi}_{1}^{0}} < m_{\tilde{a}} + m_{h}$. Figure~\ref{fig:maxino-slice-exclusions-combined} shows the projected central value (i.e. theory uncertainties are omitted) exclusions at $95\%$ confidence for varying axino masses. We observe that all the exclusions have a sharp cutoff at the kinematic threshold as expected. The lifetime of the decaying NLSP scales with $f_{a}$ and $m_{\tilde{\chi}_{1}^{0}}$ as described by eq.~\eqref{eq:lifetime_approx}. This is reflected in the projected exclusion contours in figure~\ref{fig:maxino-slice-exclusions-combined} which show that sensitivity is lower when $m_{\tilde{\chi}_{1}^{0}}$ is large and $f_{a}$ is small which leads to decays which are too prompt for a displaced vertex search. Simultaneously, the higgsino pair production cross section~\cite{Fuks:2012qx, Fuks:2013vua} decreases with larger $m_{\tilde{\chi}_{1}^{0}}$ which contributes to a narrowing of the excluded $f_{a}$ range for larger values of $m_{\tilde{\chi}_{1}^{0}}$.

\begin{figure}[htbp]
  \centering
  \subfloat[Comparison to searches which look for the axion-photon coupling~\cite{Ouellet:2018beu,Salemi:2021gck,Pandey:2024dcd,Asztalos2010,ADMX:2018gho,ADMX:2019uok,ADMX:2021nhd,Bartram:2024ovw,ADMX:2025vom, ADMX:2018ogs,Bartram:2021ysp, Crisosto:2019fcj, Lee:2020cfj,Jeong:2020cwz,CAPP:2020utb,Yoon:2022gzp,Lee:2022mnc,Kim:2022hmg,Yi:2022fmn,Yang:2023yry,Kim:2023vpo,CAPP:2024dtx,Bae:2024kmy,Adair:2022rtw,Oshima:2023csb,Devlin:2021fpq, Hoshino:2025fiz, Grenet:2021vbb,Brubaker:2016ktl,HAYSTAC:2018rwy,HAYSTAC:2020kwv,HAYSTAC:2023cam,HAYSTAC:2024jch,Heinze:2023nfb,Garcia:2024xzc,McAllister:2017lkb,Quiskamp:2022pks,Quiskamp:2023ehr,Quiskamp:2024oet, Alesini:2019ajt,Alesini:2020vny,Alesini:2022lnp,QUAX:2023gop,QUAX:2024fut,CAST:2020rlf,Ahyoune:2024klt,DePanfilis,Wuensch:1989sa,Gramolin:2020ict,TASEH:2022vvu,Arza:2021ekq,Friel:2024shg,Nishizawa:2025xka,Hagmann,Hagmann:1996qd,Thomson:2019aht,Thomson:2023moc,Ehret:2010mh,CAST:2007jps,CAST:2017uph,CAST:2024eil,Betz:2013dza,OSQAR:2015qdv,DellaValle:2015xxa,SAPPHIRES:2021vkz,SAPPHIRES:2022bqg,Kirita:2024wti,Escudero:2023vgv,Xiao:2020pra,BICEPKeck:2021sbt,Gan:2023swl,Keller:2021zbl,Chan:2021gjl,Kohri:2017ljt,Wouters:2013hua,Marsh:2017yvc,Reynolds:2019uqt,Reynes:2021bpe,Capozzi:2023xie,Liu:2023nct,Mondino:2024rif,Goldstein:2024mfp,Bolliet:2020ofj,Cyr:2024sbd,Caputo:2022mah,Calore:2020tjw,Calore:2021hhn,Buen-Abad:2020zbd,EPTA:2024gxu,Fermi-LAT:2016nkz,Meyer:2020vzy,Davies:2022wvj,Bernal:2022xyi,Sun:2023acy,Ayala:2014pea,Dolan:2022kul,Dev:2023hax,Diamond:2023cto,Jacobsen:2022swa,HESS:2013udx,Calore:2022pks,Wadekar:2021qae,Ning:2024eky,Candon:2024eah,MAGIC:2024arq,Dessert:2021bkv,Dessert:2022yqq,Benabou:2025jcv,Ivanov:2018byi,Li:2020pcn,Li:2021gxs,Li:2024zst,Foster:2020pgt,Darling:2020uyo,Battye:2021yue,Foster:2022fxn,Battye:2023oac,Perez:2016tcq,Ng:2019gch,Roach:2022lgo,Ruz:2024gkl,Fedderke:2019ajk,POLARBEAR:2023ric,POLARBEAR:2024vel,Xue:2024zjq,Noordhuis:2022ljw,Caputo:2019tms,Severino:2022nue,Vinyoles2015,Nguyen:2023czp,Jaeckel:2017tud,Hoof:2022xbe,Muller:2023vjm,Payez:2014xsa,Manzari:2024jns,Lucente:2020whw,Caputo:2021rux,Diamond:2023scc,Yuan:2020xui,Fiorillo:2025yzf,DeRocco:2022jyq,Beaufort:2023zuj,Dessert:2020lil,SPT-3G:2022ods,Wang:2023imi,Blout:2000uc,Todarello:2023hdk,Grin:2006aw,Nakayama:2022jza,Carenza:2023qxh,Todarello:2024qci,Janish:2023kvi,Pinetti:2025owq,Saha:2025any,Bessho:2022yyu,Yin:2024lla,Fong:2024qeq,Yin:2025xad,Dolan:2021rya,Foster:2021ngm,Cadamuro:2011fd,Depta:2020wmr,Langhoff:2022bij,Porras-Bedmar:2024uql}
  at a 90\% confidence level.]{\includegraphics[width=0.80\textwidth]{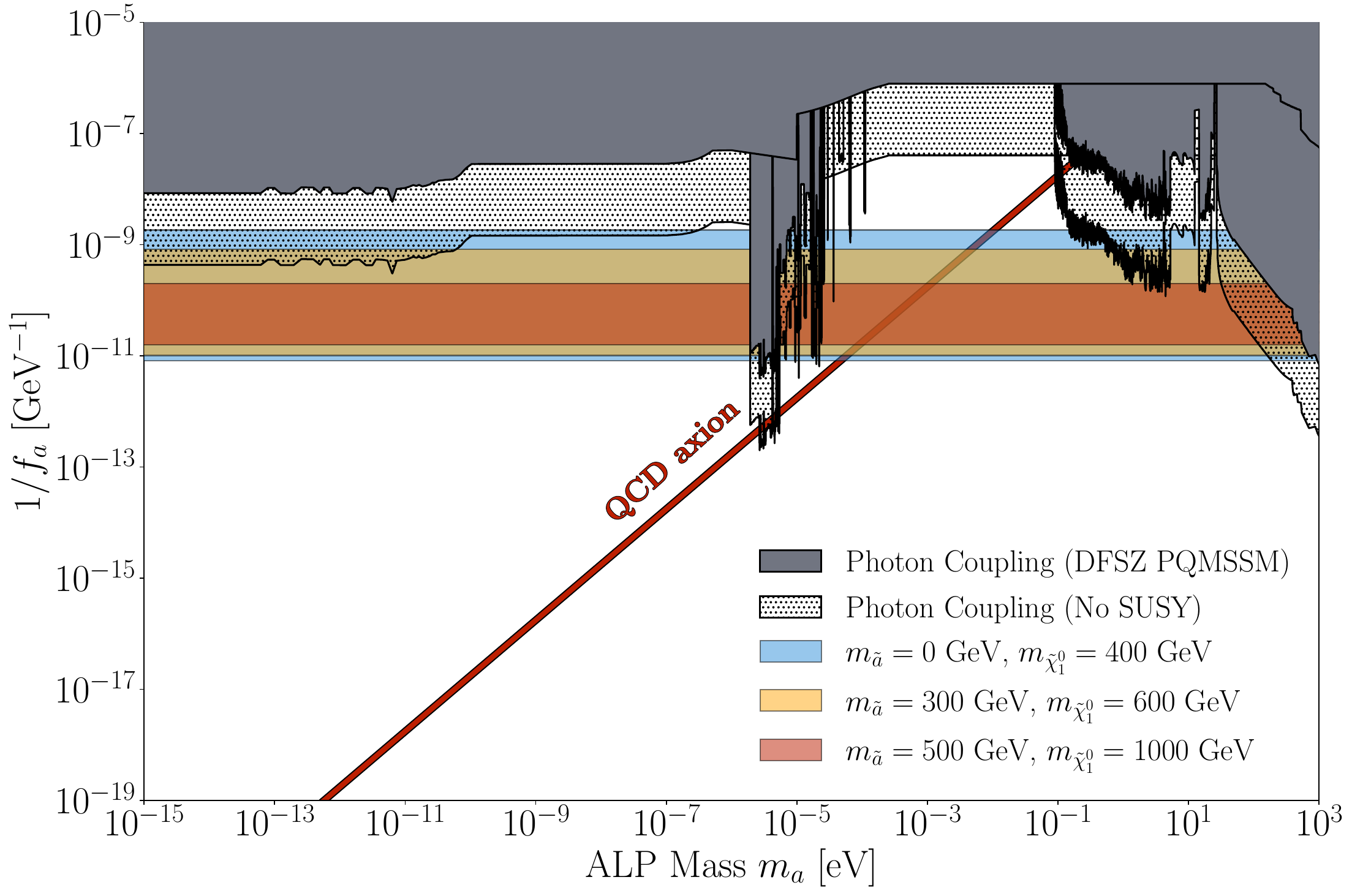}} \\
  \subfloat[Comparison to astrophysical constraints on the axion-matter coupling~\cite{neutron_star_cooling, Capozzi:2020, XENON:2019gfn, XENON:2020rca, Wei:2023rzs, Xu:2023, Lee:2022vvb} 
  and black-hole spin constraints~\cite{witte2025stepping} at a 95\% confidence level.]{\includegraphics[width=0.80\textwidth]{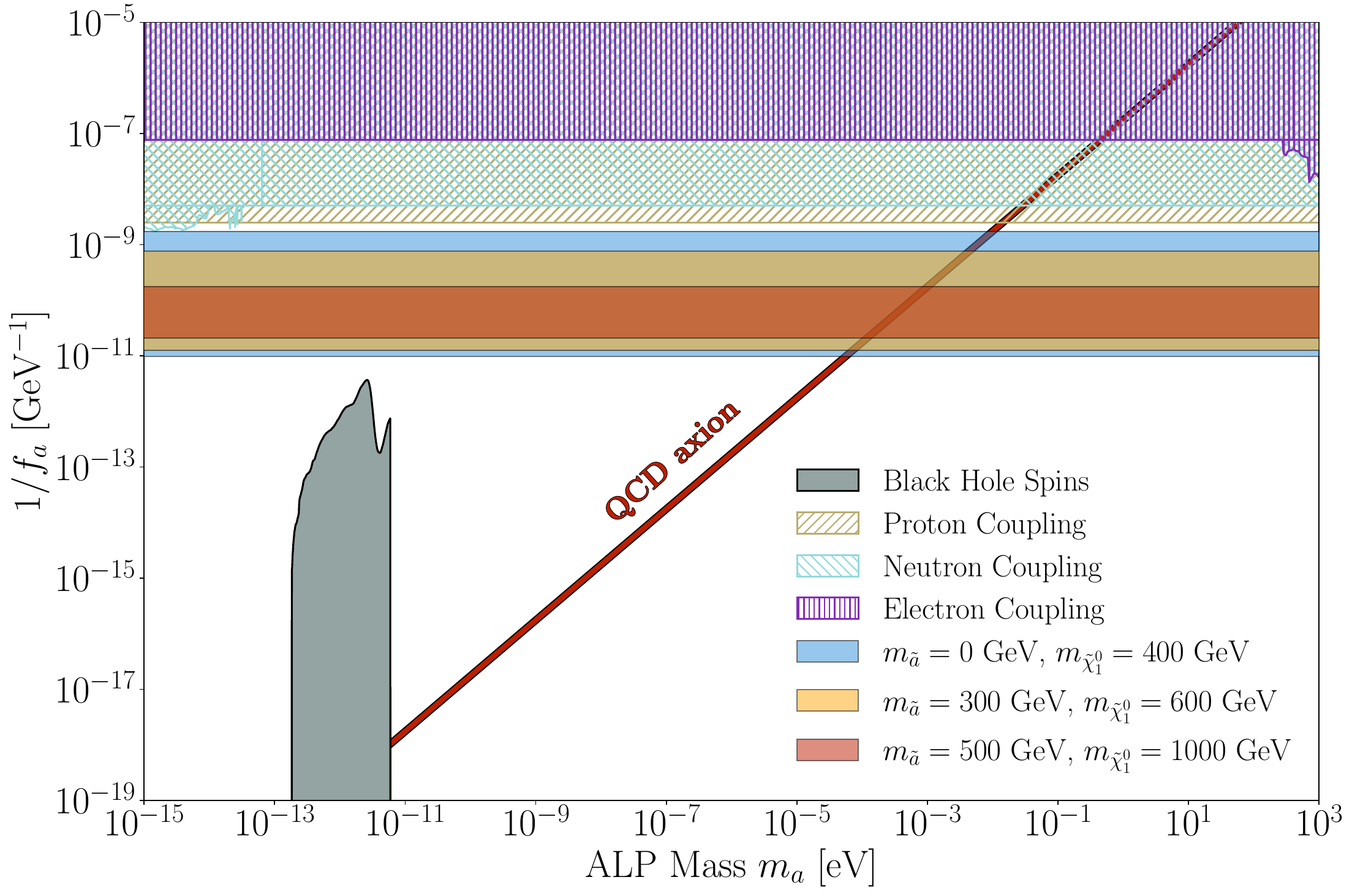}} 
  \caption{\label{fig:collider-direct-comparison} Projected sensitivity of a collider-based search for DFSZ axinos in the $f_{a}-m_{a}$ plane with different SUSY model parameters. This allows for a comparison between the sensitivities of direct detection axion experiments and a collider-based approach. The conversion between the limit on a given coupling and the limit on $f_{a}$ is calculated using the model specific couplings of the supersymmetric DFSZ axion model presented in this work. Limit plot code was modified from~\cite{AxionLimits}.}
\end{figure}

Finally we consider events which were generated for fixed points in the $m_{\tilde{\chi}_{1}^{0}}-m_{\tilde{a}}$ plane while varying $f_{a}$. The event yields for these events were calculated and an exclusion sensitivity was determined for $f_{a}$ for axions and DFSZ-like ALPs as shown in figure~\ref{fig:collider-direct-comparison}. As one would expect based on figure~\ref{fig:lifetime}, eq.~\eqref{eq:lifetime_approx}, and much of our discussion above, there is a range of excluded $f_{a}$ for each point in the $m_{\tilde{\chi}_{1}^{0}}-m_{\tilde{a}}$ plane which corresponds to decays with a lifetime that is long enough that it is readily reconstructed as a displaced vertex but not so long that the NLSP decays outside of the detector volume entirely.

\section{Joint interpretation of direct detection, astrophysical searches, and collider experiments}
\label{sec:interpretation}
The sensitivities shown in figure~\ref{fig:collider-direct-comparison} allow for a comparison between the constraints placed on QCD axions and generic ALPs by direct-detection experiments and astrophysical observation on the one hand, and the model-dependent constraints on the supersymmetric DFSZ axion and DFSZ-like ALPs discussed in this work. Figure~\ref{fig:collider-direct-comparison} shows 3 different color bands for each set of constraints determined from the different points in the $m_{\tilde{\chi}_{1}^{0}}-m_{\tilde{a}}$ plane. Given that these sensitivities are estimated using a Monte Carlo implementation of the DFSZ-PQMSSM model, the sensitivities presented here apply only to these DFSZ axion and DFSZ-like ALP models.

Figure~\ref{fig:collider-direct-comparison} shows how published constraints from the photon coupling can become less sensitive or vanish entirely when supersymmetry is introduced. Experiments which are sensitive to the non-supersymmetric DFSZ axion ($E/N = 8/3$) may have reduced sensitivity to the supersymmetric DFSZ axion which has a suppressed coupling to the photon ($E/N = 2$). Here we take $z=m_u/m_d=0.56$. Smaller $z$ is consistent with the hadron spectrum and can reduce the axion-photon coupling further.

Limits on $f_{a}$ due to black hole spins shown in figure~\ref{fig:collider-direct-comparison} are very generic as they depend on gravitational interactions and axion self-couplings~\cite{witte2025stepping}. Such couplings arise from the axion potential and can be calculated from the axion effective field theory~\cite{di2016qcd}. The black hole spin limits are thus not modified by the introduction of SUSY or the choice of UV model.

Axion-nucleon couplings arise from any model-dependent tree-level couplings to quarks which may be present and also receive a model independent contribution due to the axion coupling to gluons~\cite{di2016qcd}. DFSZ axions have tree level couplings to quarks which depend on $\tan(\beta)$ and ultimately yield a coupling to protons and neutrons which depend on $\tan(\beta)$ and have uncertainties from lattice QCD~\cite{di2016qcd, neutron_star_cooling}:
\begin{align}
C_{p} = -0.182 - 0.435 \sin^{2}(\beta) \pm 0.025,\\
C_{n} = -0.160 + 0.414 \sin^{2}(\beta) \pm 0.025
\end{align}
In the lower panel of  figure~\ref{fig:collider-direct-comparison}, $\tan(\beta)$ is taken to be 10, and 
to be conservative, the minimum coupling within lattice QCD uncertainties is used.
The axion-electron coupling is also model-dependent and may have contributions at tree level and higher order contributions due to the photon coupling~\cite{gavrilyukNewConstraintsAxion2022, choiPrecisionAxionPhysics2021}. DFSZ axions have a tree-level coupling to electrons which depends on $\tan(\beta)$ ($\propto \frac{1}{3}\cos^{2}(\beta)$)~\cite{gavrilyukNewConstraintsAxion2022, choiPrecisionAxionPhysics2021}. The electron coupling limit shown in figure~\ref{fig:collider-direct-comparison} was converted to a limit on $f_{a}$ using this model-dependent tree-level coupling.
The lower panel of figure~\ref{fig:collider-direct-comparison} demonstrates that the collider search for the higgsino and axino can complement the astrophysical constraints on the axion.

Additional constraints on other values of $f_{a}$ could come from collider searches targeting different lifetimes. ATLAS has conducted a search for a prompt NLSP decaying into a stable LSP axino, but did not attempt to interpret the results as a limit on $f_{a}$ as was done in this work~\cite{ATLAS:2021yqv}.
Whereas the search discussed here would require $f_{a}$ to be large enough to create a displaced vertex but not so large that the NLSPs decay outside of the detector, a prompt search would have sensitivity to arbitrarily small values of $f_{a}$ with a cutoff in sensitivity once the decay becomes too long-lived to be considered prompt. 
Given that the lifetime of the NLSP becomes shorter with increasing NLSP mass, we would also expect the prompt search to have more sensitivity for larger NLSP masses. 
The bands shown in figure~\ref{fig:collider-direct-comparison} are for values of $m_{\tilde{a}}$ and $m_{\tilde{\chi}_{1}^{0}}$ which are not excluded by prompt searches~\cite{ATLAS:2021yqv}, but if the constraints from prompt searches were shown for smaller $m_{\tilde{a}}$, they would generally lie above the bands for displaced vertex searches.

The misalignment mechanism prefers $f_a \sim 10^{12}$ GeV~\cite{Preskill:1982cy,Abbott:1982af,Dine:1982ah}, so it is important to probe as high in $f_a$ as possible to cover the most interesting axion parameter space. 
Higher values of $f_{a}$ would make the NLSP sufficiently long-lived to decay outside the tracker's fiducial volume and would not give rise to a detectable displaced vertex.
Still, the high-luminosity LHC will be able to probe larger values of $f_{a}$ for a similar range of higgsino masses, since the higgsino occasionally decays at a shorter distance than the typical decay length. Alternative search strategies could further extend the reach in $f_a$. The ATLAS and CMS experiments have conducted searches looking for long-lived particles decaying in their calorimeters~\cite{EXOT-2022-04} or in their muon spectrometers~\cite{EXOT-2021-32,CMS-EXO-21-008}, in the latter case even reconstructing displaced vertices using track information
from the muon spectrometer. These searches could potentially be interpreted to constrain higher values of $f_{a}$. There are also several dedicated detectors for long-lived particles which have been proposed for installation at the LHC such as CODEX-b~\cite{aielli2025codexbopeningnewwindows}, MATHUSLA~\cite{Curtin_2019}, and ANUBIS~\cite{anubiscollaboration2025anubisdetectorsensitivityneutral}. Such detectors may allow for sensitivity to even higher values of $f_{a}$ than can be achieved with ATLAS or CMS alone in the future. Disappearing track searches may also aid in pushing the sensitivity to higher $f_{a}$. A recent analysis looking for disappearing track searches assumes a higgsino LSP and is only sensitive to relatively light higgsinos and charginos~\cite{aaboudSearchLonglivedCharginos2018}. A similar analysis may be applied to the axino LSP case in which there is a disappearing chargino track which ultimately gives rise to a displaced vertex as discussed in this work. New proposed colliders like the Future Circular Collider~\cite{benediktFutureCircularColliders2020} and a $10 \ \mathrm{TeV}$ center-of-mass energy Muon Collider~\cite{accetturaMuonCollider2023} would allow for disappearing track sensitivity to even higher masses of higgsino and chargino, and may also improve sensitivity to the axino LSP at large values of $f_{a}$.

It is important to note that the collider signatures of an axino LSP may also resemble a gravitino LSP depending on the axino and gravitino masses. In SUSY models with a gravitino LSP, the NLSP decays into a SM particle and a gravitino with a lifetime:
\begin{align}
    c \tau = \left( \frac{m_{\rm NLSP}^5}{48 \pi m_{3/2}^2 M_{\mathrm{Pl}}^2 } \right)^{-1} \simeq 20~{\rm mm} \left( \frac{1~{\rm TeV}}{m_{\rm NLSP}}\right)^5 \left( \frac{m_{3/2}}{10~{\rm keV}} \right)^2.
\end{align}
If the gravitino mass is much above $\mathcal{O}(10)$ keV, the NLSP will not decay inside the LHC. Therefore, displaced vertex and missing energy signals with a non-negligible LSP mass means that the LSP is not the gravitino. 
On the other hand, a gravitino with a mass below $\mathcal{O}(10)$ keV can mimic the signals of a light axino~\cite{brandenburgSignaturesAxinosGravitinos2005, ananyevColliderConstraintsElectroweakinos2023, Redino:2015mye, fengLightGravitinosColliders2010} and it may be difficult to distinguish the light axino LSP scenario from a light gravitino LSP using the collider signature discussed in this work. Nevertheless, by measuring the NLSP lifetime, we may determine the value of $f_a$ in the axino LSP scenario, and if the complementary measurements via axion searches like those shown in figure~\ref{fig:collider-direct-comparison} and discussed above give the same value for $f_a$, it will strongly indicate the axino LSP.    

\section{Conclusion}
\label{sec:conclusion}
This work demonstrates how a collider search can provide model-dependent sensitivity to a supersymmetric DFSZ axion which is complementary to direct detection and astrophysical axion bounds. Projections based on existing collider experiments have compelling sensitivity to DFSZ-PQMSSM models with higher sensitivity to these models than direct detection searches which use the axion-photon coupling. The projected sensitivity of a collider search is also complementary to astrophysical bounds which probe the photon coupling as well as other axion couplings.

The model considered here in which there is an axino LSP with a mostly higgsino NLSP allows for the simpler implementation of the neutralino mixing matrix but more general models can be considered in the future using spectrum generators to calculate dependent model parameters for a larger variety of models. For example, a wino-like NLSP, which is predicted when the gaugino masses are given by anomaly mediation~\cite{Randall:1998uk,Giudice:1998xp} and the scalar masses are given by gravity mediation, mixes with the higgsino to obtain a coupling to electroweak bosons and the axino, leading to a similar collider signal as the higgsino-like NLSP.%
\footnote{For a large enough wino mass, however, three-body decay modes of the wino-like NLSP dominate over the two-body decay modes, which should be appropriately implemented. This is because in the effective theory after integrating out the higgsino, the wino couples to two Higgs doublets and an axino.}

Another interesting extension is small $R$-parity violation, with which the axino LSP can decay inside the detector volume to yield another displaced vertex.
The decaying axino would avoid the strong upper bound on the reheating temperature of the Universe from the overproduction of axinos~\cite{bae2012cosmology,Co:2015pka} to enable leptogenesis, which requires a reheating temperature above $10^9$ GeV~\cite{Giudice:2003jh,Buchmuller:2004nz}. 
A study searching for KSVZ axions could also be implemented, though such searches would require different signal processes as KSVZ axion models do not have tree level couplings between axions/axinos and the higgs sector as in the DFSZ axion models.

Ultimately, this work shows that there are compelling models for supersymmetric DFSZ axions which are readily probed using existing collider experiments and which provide complementary sensitivity to direct detection axion searches and astrophysical bounds. It has inspired an experimental search conducted by the ATLAS collaboration~\cite{atlascollaboration2026searchdisplaceddecayslonglived} with sensitivity to the type of models discussed here. This work also highlights the need for further modeling efforts for more general supersymmetric axion models with different mass spectra and/or $R$-parity violation. 

\acknowledgments
This work was partially supported by the DOE grant DE-SC0009924 (KH) and by the National Science Foundation under Award Number PHY-2310094 (KD, DWM, JTO, and BR). In addition, KH would like to thank the World Premier International Research Center Initiative (WPI), MEXT, Japan (Kavli IPMU), and GH and DWM would like to thank the UChicago Joint Task Force Initiative for its support.

\clearpage
\appendix
\section{Perturbative treatment of the neutralino system}
\label{sec:appendix}
The mixing between the electroweak gauginos, neutral higgsinos, and axino is implemented as a $5\times 5$ unitary matrix. In this work we consider the case where the axino remains approximately pure after mixing and is the LSP. In this work we also take $\tilde{\chi}_{1}^{0}$ and $\tilde{\chi}_{2}^{0}$ to be mostly mixtures of the two neutral higgsinos with the $\tilde{\chi}_{1}^{0}$ state being the NLSP. The allows us to write the neutralino system in the mass basis in terms of the flavor basis:

\begin{align}
\begin{pmatrix}
\tilde{\chi}_{4}^{0}\\
\tilde{\chi}_{3}^{0}\\
\tilde{\chi}_{2}^{0}\\
\tilde{\chi}_{1}^{0}\\
\tilde{a}
\end{pmatrix} \approx N\begin{pmatrix}
\tilde{B}\\
\tilde{W}^{3}\\
\tilde{H}_{d}^{0}\\
\tilde{H}_{u}^{0}\\
\tilde{a}
\end{pmatrix}\end{align}

This matrix should diagonalize the neutralino mass matrix:

\begin{align}\mathcal{M}_{\mathrm{diag}} = N^{*}\mathcal{M} N^{\dagger}\end{align}

Where the mass matrix takes the following form (with an added row/column for the axino)~\cite{choi2001analysis, bae2012cosmology}:

 \begin{align}\mathcal{M} = \begin{pmatrix}
 M_{1} & 0 & -m_{Z}c_{\beta}s_{W} & m_{Z}s_{\beta}s_{W} & 0\\
 0 & M_{2} & m_{Z}c_{\beta}c_{W} & -m_{Z}s_{\beta}c_{W} & 0\\
 -m_{Z}c_{\beta}s_{W} & m_{Z}c_{\beta}c_{W} & 0 & -\mu & y_{a}vs_{\beta}\\
 m_{Z}s_{\beta}s_{W} & -m_{Z}s_{\beta}c_{W} & -\mu & 0 & y_{a}vc_{\beta}\\
   0 & 0 & y_{a}vs_{\beta} & y_{a}vc_{\beta} & m_{\tilde{a}}
\label{eq:mass-matrix}
\end{pmatrix}\end{align}

We perturb our mixing matrix by writing:

\begin{align}N = UV\end{align}

Where we take $V$ to block-diagonalize the higgsino system $\hat{\mathcal{M}} = V\mathcal{M}V^{T}$:

\begin{align}
  V = \begin{pmatrix}
    1 & 0 & 0 & 0 & 0\\
    0 & 1 & 0 & 0 & 0\\
    0 & 0 & \frac{1}{\sqrt{2}} & - \frac{1}{\sqrt{2}} & 0\\
    0 & 0 & \frac{1}{\sqrt{2}} & \frac{1}{\sqrt{2}} & 0\\
    0 & 0 & 0 & 0 & 1\\
  \end{pmatrix}
\label{eq:zero-order-mixing}
\end{align}

and $U$ is some perturbative correction which diagonalizes the full matrix to linear order in $m_{Z}/M_{1}$ and $m_{Z}/M_{2}$ and $y_{a}$. Using \eqref{eq:mass-matrix} and \eqref{eq:zero-order-mixing}, we have:

\begin{align}
  \hat{\mathcal{M}} =
  \left(\begin{smallmatrix}
    M_{1} & 0 & \frac{m_{Z}s_{W}}{\sqrt{2}}(s_{\beta}-c_{\beta}) & \frac{m_{Z}s_{W}}{\sqrt{2}}(s_{\beta}+c_{\beta}) & 0\\  
    0 & M_{2} & \frac{m_{Z}c_{W}}{\sqrt{2}}(c_{\beta} - s_{\beta}) & -\frac{m_{Z}c_{W}}{\sqrt{2}}(c_{\beta} + s_{\beta}) & 0\\
    \frac{m_{Z}s_{W}}{\sqrt{2}}(s_{\beta} - c_{\beta}) & \frac{m_{Z}c_{W}}{\sqrt{2}}(c_{\beta} - s_{\beta}) & -\mu & 0 & \frac{y_{a}v}{\sqrt{2}}(s_{\beta} + c_{\beta})\\
    \frac{m_{Z}s_{W}}{\sqrt{2}}(s_{\beta} + c_{\beta}) & -\frac{m_{Z}c_{W}}{\sqrt{2}}(s_{\beta} + c_{\beta}) & 0 & \mu & \frac{y_{a}v}{\sqrt{2}}(c_{\beta} - s_{\beta})\\
    0 & 0 & \frac{y_{a}v}{\sqrt{2}}(s_{\beta} + c_{\beta}) & \frac{y_{a}v}{\sqrt{2}}(c_{\beta} - s_{\beta}) & m_{\tilde{a}}
  \end{smallmatrix}\right)
\end{align}

Using nondegenerate perturbation theory~\cite{bae2007mixed, bae2012cosmology}, we calculate corrections to the off-diagnal components of $U$:

\begin{align}
U_{nm}^{(1)} = \frac{\hat{\mathcal{M}}_{mn}}{\hat{\mathcal{M}}_{mm} - \hat{\mathcal{M}}_{nn}}
\end{align}

To lowest order, we find:
\begin{align}
N = \left(\begin{smallmatrix}
  1 & 0 & \frac{m_{Z}s_{W}}{2}\left(\frac{c_{\beta} - s_{\beta}}{\mu + M_{1}} + \frac{s_{\beta} + c_{\beta}}{\mu - M_{1}}\right) & \frac{m_{Z}s_{W}}{2}\left(\frac{s_{\beta} + c_{\beta}}{\mu - M_{1}} - \frac{c_{\beta} - s_{\beta}}{\mu + M_{1}}\right) & 0\\
  0 & 1 & \frac{m_{Z}c_{W}}{2}\left(\frac{s_{\beta} - c_{\beta}}{\mu + M_{2}} - \frac{s_{\beta} + c_{\beta}}{\mu - M_{2}}\right) & \frac{m_{Z}c_{W}}{2}\left(\frac{c_{\beta} - s_{\beta}}{\mu + M_{2}} - \frac{s_{\beta} + c_{\beta}}{\mu - M_{2}}\right) & 0\\
  \frac{m_{Z}s_{W}(s_{\beta} - c_{\beta})}{\sqrt{2}(M_{1} + \mu)} & \frac{m_{Z}c_{W}(c_{\beta} - s_{\beta})}{\sqrt{2}(M_{2} + \mu)} & \frac{1}{\sqrt{2}} & \frac{1}{-\sqrt{2}} & \frac{y_{a}v(s_{\beta} + c_{\beta})}{\sqrt{2}(m_{\tilde{a}} + \mu)}\\
  \frac{m_{Z}s_{W}(s_{\beta} + c_{\beta})}{\sqrt{2}(M_{1} - \mu)} & -\frac{m_{Z}c_{W}(s_{\beta} + c_{\beta})}{\sqrt{2}(M_{2} - \mu)} & \frac{1}{\sqrt{2}} & \frac{1}{\sqrt{2}} & \frac{y_{a}v (c_{\beta} - s_{\beta})}{\sqrt{2}(m_{\tilde{a}} - \mu)}\\
  0 & 0 & \frac{y_{a}v}{2}(\frac{c_{\beta} - s_{\beta}}{\mu - m_{\tilde{a}}} - \frac{s_{\beta} + c_{\beta}}{\mu + m_{\tilde{a}}}) & \frac{y_{a}v}{2}(\frac{c_{\beta} - s_{\beta}}{\mu - m_{\tilde{a}}} + \frac{s_{\beta} + c_{\beta}}{\mu + m_{\tilde{a}}}) & 1
\end{smallmatrix}\right)
\label{eq:mixing-matrix}
\end{align}

We can see from the form of the mixing matrix \eqref{eq:mixing-matrix} and the form of the axion-higgs interaction term in the superpotential \eqref{eq:superpotential} that the decays of the higgsino NLSP to an axino LSP and higgs will be suppressed by a factor of $\frac{1}{f_a}$. Moreover, mixing between the axino and higgsino states is suppressed by a factor of $\frac{1}{f_a}$ as we would expect.

We also consider the chargino mass matrix which takes the form~\cite{djouadiCharginoNeutralinoDecays2001}:
\begin{align}
\mathcal{M}_{C} = \begin{pmatrix}
M_{2} & \sqrt{2}m_{W}s_{\beta}\\
\sqrt{2}m_{W}c_{\beta} & \mu\\
\end{pmatrix}.
  \label{eq:chargino-matrix}
\end{align}
This can be diagonalized to yield eigenvalues~\cite{djouadiCharginoNeutralinoDecays2001}:
\begin{align}
m_{\tilde{\chi}_{i}^{\pm}}^{2} = \frac{1}{2}(M_{2}^{2} + \mu^{2} + 2m_{W}^{2}) \pm \frac{1}{2}(M_{2}^{2} - \mu^{2})\sqrt{1 + \frac{4m_{W}^{2}(m_{W}^{2}c_{2\beta}^{2} + M_{2}^{2} + \mu^{2} + 2M_{2}\mu s_{2\beta})}{(M_{2}^{2} - \mu^{2})^{2}}}.
\end{align}
We can expand the above in the limit where $M_{2} \gg \mu, m_{W}$. The lighter state is:

\begin{align}
m_{\tilde{\chi}_{1}^{\pm}}^{2} =& \mu^{2} + m_{W}^{2} - \frac{m_{W}^{2}(m_{W}^{2}c_{2\beta}^{2} + M_{2}^{2} + \mu^{2} + 2M_{2}\mu s_{2\beta})}{M_{2}^{2} - \mu^{2}}, \nonumber\\
m_{\tilde{\chi}_{1}^{\pm}} \simeq & \mu - \mathcal{O}\left(\frac{m_{W}^{2}}{M_{2}}\right).
\end{align}

\clearpage
\newpage

\bibliography{AxionAxinoPhysics}

@PREAMBLE{
 "\providecommand{\noopsort}[1]{}" 
 # "\providecommand{\singleletter}[1]{#1}%" 
}

@article{Ibe:2023dcu,
    author = "Ibe, Masahiro and Nakayama, Yuhei and Shirai, Satoshi",
    title = "{Precise estimate of charged Higgsino/Wino decay rate}",
    eprint = "2312.08087",
    archivePrefix = "arXiv",
    primaryClass = "hep-ph",
    reportNumber = "IPMU23-0048",
    doi = "10.1007/JHEP03(2024)012",
    journal = "JHEP",
    volume = "03",
    pages = "012",
    year = "2024"
}

@article{Martin:1997ns,
    author = "Martin, Stephen P.",
    editor = "Kane, Gordon L.",
    title = "{A Supersymmetry primer}",
    eprint = "hep-ph/9709356",
    archivePrefix = "arXiv",
    reportNumber = "FERMILAB-PUB-97-425-T",
    doi = "10.1142/9789812839657_0001",
    journal = "Adv. Ser. Direct. High Energy Phys.",
    volume = "18",
    pages = "1--98",
    year = "1998"
}

@article{Moxhay:1984am,
    author = "Moxhay, Peter and Yamamoto, Katsuji",
    title = "{{Peccei-Quinn} Symmetry Breaking by Radiative Corrections in Supergravity}",
    reportNumber = "IFP-233-UNC",
    doi = "10.1016/0370-2693(85)91655-7",
    journal = {\PLB},
    volume = "151",
    pages = "363--366",
    year = "1985"
}

@article{Murayama:1992dj,
    author = "Murayama, H. and Suzuki, H. and Yanagida, T.",
    title = "{Radiative breaking of Peccei-Quinn symmetry at the intermediate mass scale}",
    reportNumber = "TU-407",
    doi = "10.1016/0370-2693(92)91397-R",
    journal = {\PLB},
    volume = "291",
    pages = "418--425",
    year = "1992"
}

@article{tHooft:1976rip,
    author = "'t Hooft, Gerard",
    editor = "Shifman, Mikhail A.",
    title = "{Symmetry Breaking Through Bell-Jackiw Anomalies}",
    reportNumber = "PRINT-76-0254 (HARVARD)",
    doi = "10.1103/PhysRevLett.37.8",
    journal = {\PRL},
    volume = "37",
    pages = "8--11",
    year = "1976"
}

@article{Crewther:1979pi,
    author = "Crewther, R. J. and Di Vecchia, P. and Veneziano, G. and Witten, Edward",
    title = "{Chiral Estimate of the Electric Dipole Moment of the Neutron in Quantum Chromodynamics}",
    reportNumber = "CERN-TH-2735",
    doi = "10.1016/0370-2693(79)90128-X",
    journal = {\PLB},
    volume = "88",
    pages = "123",
    year = "1979",
    note = "[Erratum: \href{https://doi.org/10.1016/0370-2693(80)91025-4}{\PLB 91, 487 (1980)}]"
}

@article{Randall:1998uk,
    author = "Randall, Lisa and Sundrum, Raman",
    title = "{Out of this world supersymmetry breaking}",
    eprint = "hep-th/9810155",
    archivePrefix = "arXiv",
    reportNumber = "MIT-CTP-2788, PUPT-1815, BUHEP-98-26",
    doi = "10.1016/S0550-3213(99)00359-4",
    journal = {\NPB},
    volume = "557",
    pages = "79--118",
    year = "1999"
}

@article{Giudice:1998xp,
    author = "Giudice, Gian F. and Luty, Markus A. and Murayama, Hitoshi and Rattazzi, Riccardo",
    title = "{Gaugino mass without singlets}",
    eprint = "hep-ph/9810442",
    archivePrefix = "arXiv",
    reportNumber = "CERN-TH-98-337, LBNL-42419, LBL-42419, UCB-PTH-98-50, UMD-PP-99-037",
    doi = "10.1088/1126-6708/1998/12/027",
    journal = {\JHEP},
    volume = "12",
    pages = "027",
    year = "1998"
}

@article{Barenboim:2014kka,
    author = "Barenboim, Gabriela and Chun, Eung Jin and Jung, Sunghoon and Park, Wan Il",
    title = "{Implications of an axino LSP for naturalness}",
    eprint = "1407.1218",
    archivePrefix = "arXiv",
    primaryClass = "hep-ph",
    reportNumber = "KIAS-P14037, NSF-KITP-14-086",
    doi = "10.1103/PhysRevD.90.035020",
    journal = {\PRD},
    volume = "90",
    number = "3",
    pages = "035020",
    year = "2014"
}

@article{Co:2015pka,
    author = "Co, Raymond T. and D'Eramo, Francesco and Hall, Lawrence J. and Pappadopulo, Duccio",
    title = "{Freeze-In Dark Matter with Displaced Signatures at Colliders}",
    eprint = "1506.07532",
    archivePrefix = "arXiv",
    primaryClass = "hep-ph",
    doi = "10.1088/1475-7516/2015/12/024",
    journal = {\JCAP},
    volume = "12",
    pages = "024",
    year = "2015"
}

@article{Giudice:2003jh,
    author = "Giudice, G. F. and Notari, A. and Raidal, M. and Riotto, A. and Strumia, A.",
    title = "{Towards a complete theory of thermal leptogenesis in the SM and MSSM}",
    eprint = "hep-ph/0310123",
    archivePrefix = "arXiv",
    reportNumber = "IFUP-TH-2003-37, CERN-TH-2003-240",
    doi = "10.1016/j.nuclphysb.2004.02.019",
    journal = {\NPB},
    volume = "685",
    pages = "89--149",
    year = "2004"
}

@article{Buchmuller:2004nz,
    author = "Buchmuller, W. and Di Bari, P. and Plumacher, M.",
    title = "{Leptogenesis for pedestrians}",
    eprint = "hep-ph/0401240",
    archivePrefix = "arXiv",
    reportNumber = "DESY-03-100, UAB-FT-551, CERN-TH-2003-199",
    doi = "10.1016/j.aop.2004.02.003",
    journal = {\APNY},
    volume = "315",
    pages = "305--351",
    year = "2005"
}

@article{Co:2019jts,
    author = "Co, Raymond T. and Hall, Lawrence J. and Harigaya, Keisuke",
    title = "{Axion Kinetic Misalignment Mechanism}",
    eprint = "1910.14152",
    archivePrefix = "arXiv",
    primaryClass = "hep-ph",
    reportNumber = "LCTP-19-28",
    doi = "10.1103/PhysRevLett.124.251802",
    journal = {\PRL},
    volume = "124",
    number = "25",
    pages = "251802",
    year = "2020"
}

@article{Co:2019wyp,
    author = "Co, Raymond T. and Harigaya, Keisuke",
    title = "{Axiogenesis}",
    eprint = "1910.02080",
    archivePrefix = "arXiv",
    primaryClass = "hep-ph",
    reportNumber = "LCTP-19-27",
    doi = "10.1103/PhysRevLett.124.111602",
    journal = {\PRL},
    volume = "124",
    number = "11",
    pages = "111602",
    year = "2020"
}

@article{Caputo:2024oqc,
    author = "Caputo, Andrea and Raffelt, Georg",
    title = "{Astrophysical Axion Bounds: The 2024 Edition}",
    eprint = "2401.13728",
    archivePrefix = "arXiv",
    primaryClass = "hep-ph",
    reportNumber = "MPP-2024-13, CERN-TH-2024-013",
    doi = "10.22323/1.454.0041",
    journal = "PoS",
    volume = "COSMICWISPers",
    pages = "041",
    year = "2024"
}

@article{Maiani:1979cx,
    author = "Maiani, L.",
    title = "{All You Need to Know about the Higgs Boson}",
    journal = "Conf. Proc. C",
    volume = "7909031",
    pages = "1--52",
    year = "1979"
}

@article{Veltman:1980mj,
    author = "Veltman, M. J. G.",
    title = "{The Infrared - Ultraviolet Connection}",
    reportNumber = "Print-80-0851 (MICHIGAN)",
    journal = {\AcPPB},
    volume = "12",
    pages = "437",
    year = "1981"
}

@article{Witten:1981nf,
    author = "Witten, Edward",
    title = "{Dynamical Breaking of Supersymmetry}",
    reportNumber = "Print-81-0317 (PRINCETON)",
    doi = "10.1016/0550-3213(81)90006-7",
    journal = {\NPB},
    volume = "188",
    pages = "513",
    year = "1981"
}

@article{Kaul:1981wp,
    author = "Kaul, Romesh K.",
    title = "{Gauge Hierarchy in a Supersymmetric Model}",
    reportNumber = "TIFR-TH-81-32-REV, TIFR-TH-81-32",
    doi = "10.1016/0370-2693(82)90453-1",
    journal = {\PLB},
    volume = "109",
    pages = "19--24",
    year = "1982"
}

@article{Preskill:1982cy,
    author = "Preskill, John and Wise, Mark B. and Wilczek, Frank",
    editor = "Srednicki, M. A.",
    title = "{Cosmology of the Invisible Axion}",
    reportNumber = "HUTP-82-A048, NSF-ITP-82-103",
    doi = "10.1016/0370-2693(83)90637-8",
    journal = {\PLB},
    volume = "120",
    pages = "127--132",
    year = "1983"
}

@article{Abbott:1982af,
    author = "Abbott, L. F. and Sikivie, P.",
    editor = "Srednicki, M. A.",
    title = "{A Cosmological Bound on the Invisible Axion}",
    reportNumber = "PRINT-82-0695 (BRANDEIS)",
    doi = "10.1016/0370-2693(83)90638-X",
    journal = {\PLB},
    volume = "120",
    pages = "133--136",
    year = "1983"
}

@article{Dine:1982ah,
    author = "Dine, Michael and Fischler, Willy",
    editor = "Srednicki, M. A.",
    title = "{The Not So Harmless Axion}",
    reportNumber = "UPR-0201T",
    doi = "10.1016/0370-2693(83)90639-1",
    journal = {\PLB},
    volume = "120",
    pages = "137--141",
    year = "1983"
}

@article{Badziak:2023fsc,
    author = "Badziak, Marcin and Harigaya, Keisuke",
    title = "{Naturally astrophobic QCD axion}",
    eprint = "2301.09647",
    archivePrefix = "arXiv",
    primaryClass = "hep-ph",
    doi = "10.1007/JHEP06(2023)014",
    journal = {\JHEP},
    volume = "06",
    pages = "014",
    year = "2023"
}

@article{Peccei:1977hh,
    author = "Peccei, R. D. and Quinn, Helen R.",
    title = "{CP Conservation in the Presence of Instantons}",
    reportNumber = "ITP-568-STANFORD",
    doi = "10.1103/PhysRevLett.38.1440",
    journal = {\PRL},
    volume = "38",
    pages = "1440--1443",
    year = "1977"
}

@article{Peccei:1977ur,
    author = "Peccei, R. D. and Quinn, Helen R.",
    title = "{Constraints Imposed by CP Conservation in the Presence of Instantons}",
    reportNumber = "ITP-572-STANFORD",
    doi = "10.1103/PhysRevD.16.1791",
    journal = {\PRD},
    volume = "16",
    pages = "1791--1797",
    year = "1977"
}

@article{Redino:2015mye,
  title = {{Exploring the hadronic axion window via delayed neutralino decay to axinos at the LHC}},
  author = {Redino, C. S. and Wackeroth, D.},
  journal = {\PRD},
  volume = {93},
  issue = {7},
  pages = {075022},
  numpages = {20},
  year = {2016},
  month = {Apr},
  publisher = {American Physical Society},
  doi = {10.1103/PhysRevD.93.075022},
  url = {https://link.aps.org/doi/10.1103/PhysRevD.93.075022}
}

@article{KIM1984150,
title = {{The $\mu$-problem and the strong $CP$-problem}},
journal = {\PLB},
volume = {138},
number = {1},
pages = {150-154},
year = {1984},
issn = {0370-2693},
doi = {10.1016/0370-2693(84)91890-2},
url = {https://www.sciencedirect.com/science/article/pii/0370269384918902},
author = {Jihn E. Kim and H.P. Nilles}
}

@article{choi2001analysis,
  title="{Analysis of the neutralino system in supersymmetric theories}",
  author={Choi, SY and Kalinowski, J and Moortgat--Pick, G and Zerwas, PM},
  journal={\EPJC},
  volume={22},
  number={3},
  pages={563--579},
  year={2001},
  publisher={Springer},
  eprint = "hep-ph/0108117",
  archivePrefix = "arXiv",
  primaryClass = "",
  doi={10.1007/s100520100808}
}

@article{bae2012cosmology,
  title="{Cosmology of the DFSZ axino}",
  author={Bae, Kyu Jung and Chun, Eung Jin and Im, Sang Hui},
  journal={\JCAP},
  volume={2012},
  number={03},
  pages={013},
  year={2012},
  publisher={IOP Publishing},
  eprint = "1111.5962",
  archivePrefix = "arXiv",
  primaryClass = "hep-ph",
  doi = {10.1088/1475-7516/2012/03/013}
}

@article{bae2007mixed,
  title="{Mixed bino--wino--higgsino dark matter in gauge messenger models}",
  author={Bae, Kyu Jung and Derm{\'\i}{\v{s}}ek, Radovan and Do Kim, Hyung and Kim, Ian-Woo},
  journal={\JCAP},
  volume={2007},
  number={08},
  pages={014},
  year={2007},
  publisher={IOP Publishing},
  doi={10.1088/1475-7516/2007/08/014},
  eprint = "hep-ph/0702041",
  archivePrefix = "arXiv",
  primaryClass = ""
}

@article{raffelt1990astrophysical,
  title="{Astrophysical methods to constrain axions and other novel particle phenomena}",
  author={Raffelt, Georg G},
  journal={\PRep},
  volume={198},
  number={1-2},
  pages={1--113},
  year={1990},
  publisher={Elsevier},
  doi = {10.1016/0370-1573(90)90054-6},
}

@article{liu2022broadband,
    title="{Broadband solenoidal haloscope for terahertz axion detection}",
    author={Liu, Jesse and Dona, Kristin and Hoshino, Gabe and Knirck, Stefan and Kurinsky, Noah and Malaker, Matthew and Miller, David W and Sonnenschein, Andrew and Awida, Mohamed H and Barry, Peter S and others},
    journal={\PRL},
    volume={128},
    number={13},
    pages={131801},
    year={2022},
    publisher={APS},
    eprint = "2111.12103",
    archivePrefix = "arXiv",
    primaryClass = "physics.ins-det",
    doi = "10.1103/PhysRevLett.128.131801"
}

@article{abel2020measurement,
    title={Measurement of the permanent electric dipole moment of the neutron},
    author={Abel, Christopher and Afach, Samer and Ayres, Nicholas J and Baker, Colin A and Ban, Gilles and Bison, Georg and Bodek, Kazimierz and Bondar, Vira and Burghoff, Martin and Chanel, E and others},
    eprint = "2001.11966",
    archivePrefix = "arXiv",
    primaryClass = "hep-ex",
    doi = "10.1103/PhysRevLett.124.081803",
    journal = {\PRL},
    volume = "124",
    number = "8",
    pages = "081803",
    year = "2020",
    publisher={APS}
}

@article{bae2017prospects,
  title={{Prospects for axion detection in natural SUSY with mixed axion-higgsino dark matter: back to invisible?}},
  author={Bae, Kyu Jung and Baer, Howard and Serce, Hasan},
  journal={\JCAP},
  volume={2017},
  number={06},
  pages={024},
  year={2017},
  publisher={IOP Publishing},
  doi = "10.1088/1475-7516/2017/06/024",
  eprint = "1705.01134",
  archivePrefix = "arXiv",
  primaryClass = "hep-ph"
}

@article{bajjali2023first,
    author={Bajjali, Fayez and Dornbusch, Sven and Ekmed{\v{z}}i{\'c}, Marko and Horns, Dieter and Kasemann, Christoph and Lobanov, Andrei and Nguyen, Le Hoang and Tluczykont, Martin and Tuccari, Gino and Ulrichs, Johannes and others},
    title = "{First results from BRASS-p broadband searches for hidden photon dark matter}",
    eprint = "2306.05934",
    archivePrefix = "arXiv",
    primaryClass = "hep-ex",
    doi = "10.1088/1475-7516/2023/08/077",
    journal = {\JCAP},
    volume = "08",
    pages = "077",
    year = "2023"
}

@article{silva-feaver2017,
  author={Silva-Feaver, Maximiliano and Chaudhuri, Saptarshi and Cho, Hsaio-Mei and Dawson, Carl and Graham, Peter and Irwin, Kent and Kuenstner, Stephen and Li, Dale and Mardon, Jeremy and Moseley, Harvey and Mule, Richard and Phipps, Arran and Rajendran, Surjeet and Steffen, Zach and Young, Betty},
  journal={\IEEETAS}, 
  title="{Design Overview of DM Radio Pathfinder Experiment}", 
  year={2017},
  volume={27},
  number={4},
  pages={1-4},
  doi={10.1109/TASC.2016.2631425},
  eprint = "1610.09344",
  archivePrefix = "arXiv",
  primaryClass = "astro-ph.IM"
}

@article{araz2021simplified,
    author = "Araz, Jack Y. and Fuks, Benjamin and Polykratis, Georgios",
    title = "{Simplified fast detector simulation in MADANALYSIS 5}",
    eprint = "2006.09387",
    archivePrefix = "arXiv",
    primaryClass = "hep-ph",
    doi = "10.1140/epjc/s10052-021-09052-5",
    journal={\EPJC},
    publisher={Springer},
    volume = "81",
    number = "4",
    pages = "329",
    year = "2021"
}

@article{ATLAS:2017tny,
    author = "Aaboud, Morad and others",
    collaboration = "ATLAS",
    title = "{Search for long-lived, massive particles in events with displaced vertices and missing transverse momentum in $\sqrt{s}$ = 13 TeV $pp$ collisions with the ATLAS detector}",
    eprint = "1710.04901",
    archivePrefix = "arXiv",
    primaryClass = "hep-ex",
    reportNumber = "CERN-EP-2017-202",
    doi = "10.1103/PhysRevD.97.052012",
    journal = {\PRD},
    volume = "97",
    number = "5",
    pages = "052012",
    year = "2018"
}

@misc{DVN/31JVGJ_2021,
    author = {Utsch, Manuel and Goodsell, Mark},
    publisher = {Open Data @ UCLouvain},
    title = {{Implementation of a search for a displaced vertices with oppositely-charged leptons (32.8 fb-1; 13 TeV; ATLAS-SUSY-2017-04)}},
    year = {2021},
    version = {V1},
    doi = {10.14428/DVN/31JVGJ},
    url = {https://doi.org/10.14428/DVN/31JVGJ}
}

@article{Fuks:2012qx,
    author = "Fuks, Benjamin and Klasen, Michael and Lamprea, David R. and Rothering, Marcel",
    title = "{Gaugino production in proton-proton collisions at a center-of-mass energy of 8 TeV}",
    eprint = "1207.2159",
    archivePrefix = "arXiv",
    primaryClass = "hep-ph",
    reportNumber = "IPHC-PHENO-12-07, MS-TP-12-05",
    doi = "10.1007/JHEP10(2012)081",
    journal = {\JHEP},
    volume = "10",
    pages = "081",
    year = "2012"
}

@article{Fuks:2013vua,
    author         = "Fuks, Benjamin and Klasen, Michael and Lamprea, David R.
                    and Rothering, Marcel",
    title          = "{Precision predictions for electroweak superpartner
                    production at hadron colliders with {\sc Resummino}}",
    journal        = {\EPJC},
    volume         = "73",
    pages          = "2480",
    doi            = "10.1140/epjc/s10052-013-2480-0",
    year           = "2013",
    eprint         = "1304.0790",
    archivePrefix  = "arXiv",
    primaryClass   = "hep-ph",
    reportNumber   = "CERN-PH-TH-2013-064, IPHC-PHENO-13-02, MS-TP-13-06",
    SLACcitation   = "%%CITATION = ARXIV:1304.0790;%%"
}

@article{beylin2008diagonalization,
    author = "Beylin, V. A. and Kuksa, V. I. and Pasechnik, R. S. and Vereshkov, G. M.",
    title = "{Diagonalization of the neutralino mass matrix and boson-neutralino interaction}",
    eprint = "hep-ph/0702148",
    archivePrefix = "arXiv",
    doi = "10.1140/epjc/s10052-008-0660-0",
    journal = {\EPJC},
    volume = "56",
    pages = "395--405",
    year = "2008"
}

@article{witte2025stepping,
    author = "Witte, Samuel J. and Mummery, Andrew",
    title = "{Stepping up superradiance constraints on axions}",
    eprint = "2412.03655",
    archivePrefix = "arXiv",
    primaryClass = "hep-ph",
    doi = "10.1103/PhysRevD.111.083044",
    journal = {\PRD},
    volume = "111",
    number = "8",
    pages = "083044",
    year = "2025"
}

@article{Capozzi:2020,
  title = {Axion and neutrino bounds improved with new calibrations of the tip of the red-giant branch using geometric distance determinations},
  author = {Capozzi, Francesco and Raffelt, Georg},
  journal = {\PRD},
  volume = {102},
  issue = {8},
  pages = {083007},
  numpages = {14},
  year = {2020},
  month = {Oct},
  publisher = {American Physical Society},
  doi = {10.1103/PhysRevD.102.083007},
  url = {https://link.aps.org/doi/10.1103/PhysRevD.102.083007}
}

@article{di2016qcd,
    author = "Grilli di Cortona, Giovanni and Hardy, Edward and Pardo Vega, Javier and Villadoro, Giovanni",
    title = "{The QCD axion, precisely}",
    eprint = "1511.02867",
    archivePrefix = "arXiv",
    primaryClass = "hep-ph",
    doi = "10.1007/JHEP01(2016)034",
    journal = {\JHEP},
    volume = "01",
    pages = "034",
    year = "2016"
}

@article{neutron_star_cooling,
    author = "Buschmann, Malte and Dessert, Christopher and Foster, Joshua W. and Long, Andrew J. and Safdi, Benjamin R.",
    title = "{Upper Limit on the QCD Axion Mass from Isolated Neutron Star Cooling}",
    eprint = "2111.09892",
    archivePrefix = "arXiv",
    primaryClass = "hep-ph",
    doi = "10.1103/PhysRevLett.128.091102",
    journal = {\PRL},
    volume = "128",
    number = "9",
    pages = "091102",
    year = "2022",
    month = {Mar},
    publisher = {American Physical Society},
}

@article{gavrilyukNewConstraintsAxion2022,
  title={New Constraints on the Axion--Electron Coupling Constant for Solar Axions},
  author={Gavrilyuk, Yu M and Gangapshev, AN and Derbin, Aleksandr Vladimirovich and Drachnev, Ilya Sergeevich and Kazalov, Vladimir Vladimirovich and Kuzminov, Valerii Vasil'evich and Mikulich, MS and Muratova, Valentina Nikolaevna and Tekueva, Djamilya Anuarovna and Unzhakov, Evgeniy Vadimovich and others},
  journal={JETP Letters},
  volume={116},
  number={1},
  pages={11--17},
  year={2022},
  publisher={Springer},
  doi = {10.1134/S0021364022601075},
  url = {https://link.springer.com/10.1134/S0021364022601075},
  }

@article{choiPrecisionAxionPhysics2021,
  title = {Precision Axion Physics with Running Axion Couplings},
  author = {Choi, Kiwoon and Im, Sang Hui and Kim, Hee Jung and Seong, Hyeonseok},
  year = {2021},
  month = {aug},
  journal = {\JHEP},
  volume = {2021},
  number = {8},
  pages = {58},
  issn = {1029-8479},
  doi = {10.1007/JHEP08(2021)058},
  url = {https://link.springer.com/10.1007/JHEP08(2021)058},
  urldate = {2025-09-17},
  langid = {english}
}

@article{brandenburgSignaturesAxinosGravitinos2005,
  title = {Signatures of Axinos and Gravitinos at Colliders},
  author = {Brandenburg, A. and Covi, L. and Hamaguchi, K. and Roszkowski, L. and Steffen, F.D.},
  year = {2005},
  month = {jun},
  journal = {\PLB},
  volume = {617},
  number = {1-2},
  pages = {99--111},
  issn = {03702693},
  doi = {10.1016/j.physletb.2005.04.072},
  url = {https://linkinghub.elsevier.com/retrieve/pii/S0370269305006015},
  urldate = {2025-09-17},
  copyright = {https://www.elsevier.com/tdm/userlicense/1.0/},
  langid = {english},
}

@article{ananyevColliderConstraintsElectroweakinos2023,
  title = {Collider Constraints on Electroweakinos in the Presence of a Light Gravitino},
  author = {Ananyev, Viktor and Bal{\'a}zs, Csaba and Beniwal, Ankit and Braseth, Lasse Lorentz and Buckley, Andy and Butterworth, Jonathan and Chang, Christopher and Danninger, Matthias and Fowlie, Andrew and Gonzalo, Tom{\'a}s E. and Kvellestad, Anders and Mahmoudi, Farvah and Martinez, Gregory D. and Prim, Markus T. and Procter, Tomasz and Raklev, Are and Scott, Pat and St{\"o}cker, Patrick and Van Den Abeele, Jeriek and White, Martin and Zhang, Yang and {GAMBIT Collaboration}},
  year = {2023},
  month = {jun},
  journal = {\EPJC},
  volume = {83},
  number = {6},
  pages = {493},
  issn = {1434-6052},
  doi = {10.1140/epjc/s10052-023-11574-z},
  url = {https://link.springer.com/10.1140/epjc/s10052-023-11574-z},
  urldate = {2025-09-17},
  langid = {english},
}

@article{fengLightGravitinosColliders2010,
  title = {Light Gravitinos at Colliders and Implications for Cosmology},
  author = {Feng, Jonathan L. and Kamionkowski, Marc and Lee, Samuel K.},
  year = {2010},
  month = {jul},
  journal = {\PRD},
  volume = {82},
  number = {1},
  pages = {015012},
  issn = {1550-7998, 1550-2368},
  doi = {10.1103/PhysRevD.82.015012},
  url = {https://link.aps.org/doi/10.1103/PhysRevD.82.015012},
  urldate = {2025-09-18},
  copyright = {http://link.aps.org/licenses/aps-default-license},
  langid = {english},
}

@article{zhitnitsky_possible_1980,
	title = {On {Possible} {Suppression} of the {Axion} {Hadron} {Interactions}.},
	volume = {31},
	journal = {\SJNP},
	author = {Zhitnitsky, A. R.},
	year = {1980},
	pages = {260}
}

@article{dine_simple_1981,
	title = {A simple solution to the strong {CP} problem with a harmless axion},
	volume = {104},
	issn = {0370-2693},
	url = {https://www.sciencedirect.com/science/article/pii/0370269381905906},
	doi = {10.1016/0370-2693(81)90590-6},
	number = {3},
	journal = {\PLB},
	author = {Dine, Michael and Fischler, Willy and Srednicki, Mark},
	year = {1981},
	pages = {199--202},
}

@article{kim_weak-interaction_1979,
	title = "{Weak-Interaction Singlet and Strong $\mathrm{CP}$ Invariance}",
	volume = {43},
	url = {https://link.aps.org/doi/10.1103/PhysRevLett.43.103},
	doi = {10.1103/PhysRevLett.43.103},
	number = {2},
	journal = {\PRL},
	author = {Kim, Jihn E.},
	month = 7,
	year = {1979},
	pages = {103--107},
}

@article{shifman_can_1980,
	title = {Can confinement ensure natural {CP} invariance of strong interactions?},
	volume = {166},
	issn = {0550-3213},
	url = {https://www.sciencedirect.com/science/article/pii/0550321380902096},
	doi = {10.1016/0550-3213(80)90209-6},
	number = {3},
	journal = {\NPB},
	author = {Shifman, M. A. and Vainshtein, A. I. and Zakharov, V. I.},
	year = {1980},
	pages = {493--506},
}

@article{ATLAS:MainPaper,
    author         = "{ATLAS Collaboration}",
    title          = "{The ATLAS Experiment at the CERN Large Hadron Collider}",
    journal        = "JINST",
    volume         = "3",
    year           = "2008",
    pages          = "S08003",
    doi            = "10.1088/1748-0221/3/08/S08003",
    primaryClass   = "hep-ex",
}

@article{CMS:2008xjf,
    author         = "{CMS Collaboration}",
    title          = "{The CMS Experiment at the CERN LHC}",
    journal        = "JINST",
    volume         = "3",
    year           = "2008",
    pages          = "S08004",
    doi            = "10.1088/1748-0221/3/08/S08004",
}

@article{Madgrah:2014,
    author = "Alwall, J. and Frederix, R. and Frixione, S. and Hirschi, V. and Maltoni, F. and Mattelaer, O. and Shao, H. -S. and Stelzer, T. and Torrielli, P. and Zaro, M.",
    title = "{The automated computation of tree-level and next-to-leading order differential cross sections, and their matching to parton shower simulations}",
    eprint = "1405.0301",
    archivePrefix = "arXiv",
    primaryClass = "hep-ph",
    reportNumber = "CERN-PH-TH-2014-064, CP3-14-18, LPN14-066, MCNET-14-09, ZU-TH-14-14",
    doi = "10.1007/JHEP07(2014)079",
    journal = {\JHEP},
    volume = "07",
    pages = "079",
    year = "2014"
}

@article{Pythia8.3:2022,
    author = "Bierlich, Christian and others",
    title = "{A comprehensive guide to the physics and usage of PYTHIA 8.3}",
    eprint = "2203.11601",
    archivePrefix = "arXiv",
    primaryClass = "hep-ph",
    reportNumber = "LU-TP 22-16, MCNET-22-04, FERMILAB-PUB-22-227-SCD",
    doi = "10.21468/SciPostPhysCodeb.8",
    journal = "SciPost Phys. Codeb.",
    volume = "2022",
    pages = "8",
    year = "2022"
}

@article{MadAnalysis5:2012,
    author = "Conte, Eric and Fuks, Benjamin and Serret, Guillaume",
    title = "{MadAnalysis 5, A User-Friendly Framework for Collider Phenomenology}",
    eprint = "1206.1599",
    archivePrefix = "arXiv",
    primaryClass = "hep-ph",
    reportNumber = "IPHC-PHENO-06",
    doi = "10.1016/j.cpc.2012.09.009",
    journal = {\CPC},
    volume = "184",
    pages = "222--256",
    year = "2013"
}

@article{SARAH:2012,
    author = "Staub, Florian and Ohl, Thorsten and Porod, Werner and Speckner, Christian",
    title = "{A Tool Box for Implementing Supersymmetric Models}",
    eprint = "1109.5147",
    archivePrefix = "arXiv",
    primaryClass = "hep-ph",
    reportNumber = "FR-PHENO-2011-017",
    doi = "10.1016/j.cpc.2012.04.013",
    journal = {\CPC},
    volume = "183",
    pages = "2165--2206",
    year = "2012"
}

@article{SARAH4:2014,
    author = "Staub, Florian",
    title = "{\texttt{SARAH 4}: A tool for (not only SUSY) model builders}",
    eprint = "1309.7223",
    archivePrefix = "arXiv",
    primaryClass = "hep-ph",
    reportNumber = "BONN-TH-2013-17",
    doi = "10.1016/j.cpc.2014.02.018",
    journal = {\CPC},
    volume = "185",
    pages = "1773--1790",
    year = "2014"
}

@article{UFO:2011,
    author = "Degrande, Celine and Duhr, Claude and Fuks, Benjamin and Grellscheid, David and Mattelaer, Olivier and Reiter, Thomas",
    title = "{UFO - The Universal FeynRules Output}",
    eprint = "1108.2040",
    archivePrefix = "arXiv",
    primaryClass = "hep-ph",
    reportNumber = "CP3-11-25, IPHC-PHENO-11-04, IPPP-11-39, DCPT-11-78, MPP-2011-68",
    doi = "10.1016/j.cpc.2012.01.022",
    journal = {\CPC},
    volume = "183",
    pages = "1201--1214",
    year = "2012"
}

@article{FastJet:2012,
    author = "Cacciari, Matteo and Salam, Gavin P. and Soyez, Gregory",
    title = "{FastJet User Manual}",
    eprint = "1111.6097",
    archivePrefix = "arXiv",
    primaryClass = "hep-ph",
    reportNumber = "CERN-PH-TH-2011-297",
    doi = "10.1140/epjc/s10052-012-1896-2",
    journal = {\EPJC},
    volume = "72",
    pages = "1896",
    year = "2012"
}

@article{PartonDistributions:2012,
    author = "Ball, Richard D. and others",
    title = "{Parton distributions with LHC data}",
    eprint = "1207.1303",
    archivePrefix = "arXiv",
    primaryClass = "hep-ph",
    reportNumber = "EDINBURGH-2012-08, IFUM-FT-997, FR-PHENO-2012-014, RWTH-TTK-12-25, CERN-PH-TH-2012-037, SFB-CPP-12-47",
    doi = "10.1016/j.nuclphysb.2012.10.003",
    journal = {\NPB},
    volume = "867",
    pages = "244--289",
    year = "2013"
}

@article{MadSpin:2012,
    author = "Artoisenet, Pierre and Frederix, Rikkert and Mattelaer, Olivier and Rietkerk, Robbert",
    title = "{Automatic spin-entangled decays of heavy resonances in Monte Carlo simulations}",
    eprint = "1212.3460",
    archivePrefix = "arXiv",
    primaryClass = "hep-ph",
    reportNumber = "NIKHEF-2012-021, CERN-PH-TH-2012-329",
    doi = "10.1007/JHEP03(2013)015",
    journal = {\JHEP},
    volume = "03",
    pages = "015",
    year = "2013"
}

@misc{AtlasA14:2014,
      author = "{The ATLAS Collaboration}",
      title         = "{ATLAS Pythia 8 tunes to 7 TeV data}",
      institution   = "CERN",
      reportNumber  = "ATL-PHYS-PUB-2014-021",
      address       = "Geneva",
      year          = "2014",
      url           = "https://cds.cern.ch/record/1966419",
    note = "{ATL-PHYS-PUB-2014-021}"
}

@misc{CERN_4,
      author        = "de Florian, D. and others",
      collaboration = "LHC Higgs Cross Section Working Group",
      title         = "{Handbook of LHC Higgs Cross Sections: 4. Deciphering the
                       Nature of the Higgs Sector}",
      archivePrefix = "arXiv",
      eprint        = "1610.07922",
      reportNumber  = "CERN-2017-002-M, CERN-2017-002",
      address       = "Geneva",
      year          = "2017",
      url           = "https://cds.cern.ch/record/2227475",
      doi           = "10.23731/CYRM-2017-002",
}

@article{ATLAS:2021yqv,
    author = "Aad, Georges and others",
    collaboration = "ATLAS",
    title = "{Search for charginos and neutralinos in final states with two boosted hadronically decaying bosons and missing transverse momentum in $pp$ collisions at $\sqrt {s}$ = 13{\,}{\,}TeV with the ATLAS detector}",
    eprint = "2108.07586",
    archivePrefix = "arXiv",
    primaryClass = "hep-ex",
    reportNumber = "CERN-EP-2021-127",
    doi = "10.1103/PhysRevD.104.112010",
    journal = {\PRD},
    volume = "104",
    number = "11",
    pages = "112010",
    year = "2021"
}

@article{djouadiCharginoNeutralinoDecays2001,
  title = {Chargino and Neutralino Decays Revisited},
  author = {Djouadi, A. and Mambrini, Y. and M{\"u}hlleitner, M.},
  year = {2001},
  month = {may},
  journal = {\EPJC},
  volume = {20},
  number = {3},
  pages = {563--584},
  issn = {1434-6044, 1434-6052},
  doi = {10.1007/s100520100679},
  url = {http://link.springer.com/10.1007/s100520100679},
  urldate = {2025-10-06},
}

@misc{AxionLimits,
  author       = {Ciaran O'Hare},
  title        = {cajohare/AxionLimits: AxionLimits},
  month        = jul,
  year         = 2020,
  publisher    = {Zenodo},
  version      = {v1.0},
  doi          = {10.5281/zenodo.3932430},
  howpublished = {\url{https://cajohare.github.io/AxionLimits/}}
}

@article{XENON:2019gfn,
    author = "Aprile, E. and others",
    collaboration = "XENON",
    title = "{Light Dark Matter Search with Ionization Signals in XENON1T}",
    eprint = "1907.11485",
    archivePrefix = "arXiv",
    primaryClass = "hep-ex",
    doi = "10.1103/PhysRevLett.123.251801",
    journal = {\PRL},
    volume = "123",
    number = "25",
    pages = "251801",
    year = "2019"
}

@article{XENON:2020rca,
    author = "Aprile, E. and others",
    collaboration = "XENON",
    title = "{Excess electronic recoil events in XENON1T}",
    eprint = "2006.09721",
    archivePrefix = "arXiv",
    primaryClass = "hep-ex",
    doi = "10.1103/PhysRevD.102.072004",
    journal = {\PRD},
    volume = "102",
    number = "7",
    pages = "072004",
    year = "2020"
}

@article{Wei:2023rzs,
   title={Dark matter search with a resonantly-coupled hybrid spin system},
   volume={88},
   ISSN={1361-6633},
   DOI={10.1088/1361-6633/adca52},
   number={5},
   journal={\RPP},
   publisher={IOP Publishing},
   author="Wei, Kai and others",
   year={2025},
   month=apr, pages={057801} }

@article{Xu:2023,
    author = "Xu, Zitong and others",
    title = "{Constraining ultralight dark matter through an accelerated resonant search}",
    eprint = "2309.16600",
    archivePrefix = "arXiv",
    primaryClass = "hep-ph",
    doi = "10.1038/s42005-024-01713-7",
    journal = {\CommP},
    volume = "7",
    number = "1",
    pages = "226",
    year = "2024"
}

@article{Lee:2022vvb,
  title = {Laboratory Constraints on the Neutron-Spin Coupling of feV-Scale Axions},
  author = {Lee, Junyi and Lisanti, Mariangela and Terrano, William A. and Romalis, Michael},
  journal = {\PRX},
  volume = {13},
  issue = {1},
  pages = {011050},
  numpages = {25},
  year = {2023},
  month = {Mar},
  publisher = {American Physical Society},
  doi = {10.1103/PhysRevX.13.011050},
  url = {https://link.aps.org/doi/10.1103/PhysRevX.13.011050}
}

@article{Ouellet:2018beu,
    author = "Ouellet, Jonathan L. and others",
    title = "{First Results from ABRACADABRA-10 cm: A Search for Sub-$\mu$eV Axion Dark Matter}",
    eprint = "1810.12257",
    archivePrefix = "arXiv",
    primaryClass = "hep-ex",
    doi = "10.1103/PhysRevLett.122.121802",
    journal = {\PRL},
    volume = "122",
    number = "12",
    pages = "121802",
    year = "2019"
}

@article{Salemi:2021gck,
    author = "Salemi, Chiara P. and others",
    title = "{Search for Low-Mass Axion Dark Matter with ABRACADABRA-10~cm}",
    eprint = "2102.06722",
    archivePrefix = "arXiv",
    primaryClass = "hep-ex",
    doi = "10.1103/PhysRevLett.127.081801",
    journal = {\PRL},
    volume = "127",
    number = "8",
    pages = "081801",
    year = "2021"
}

@article{Pandey:2024dcd,
    author = "Pandey, Swadha and Hall, Evan D. and Evans, Matthew",
    title = "{First Results from the Axion Dark-Matter Birefringent Cavity (ADBC) Experiment}",
    eprint = "2404.12517",
    archivePrefix = "arXiv",
    primaryClass = "hep-ex",
    doi = "10.1103/PhysRevLett.133.111003",
    journal = {\PRL},
    volume = "133",
    number = "11",
    pages = "111003",
    year = "2024"
}

@ARTICLE{Asztalos2010,
       author = {{Asztalos}, S.~J. and {Carosi}, G. and {Hagmann}, C. and {Kinion}, D. and {van Bibber}, K. and {Hotz}, M. and {Rosenberg}, L.~J. and {Rybka}, G. and {Hoskins}, J. and {Hwang}, J. and {Sikivie}, P. and {Tanner}, D.~B. and {Bradley}, R. and {Clarke}, J. and {ADMX Collaboration}},
        title = "{SQUID-Based Microwave Cavity Search for Dark-Matter Axions}",
      journal = {\PRL},
         year = 2010,
        month = jan,
       volume = {104},
       number = {4},
          eid = {041301},
        pages = {041301},
          doi = {10.1103/PhysRevLett.104.041301},
archivePrefix = {arXiv},
       eprint = {0910.5914},
 primaryClass = {astro-ph.CO},
       adsurl = {https://ui.adsabs.harvard.edu/abs/2010PhRvL.104d1301A}
}

@article{ADMX:2018gho,
    author = "Du, N. and others",
    collaboration = "ADMX",
    title = "{A Search for Invisible Axion Dark Matter with the Axion Dark Matter Experiment}",
    eprint = "1804.05750",
    archivePrefix = "arXiv",
    primaryClass = "hep-ex",
    reportNumber = "FERMILAB-PUB-18-101-AD-AE",
    doi = "10.1103/PhysRevLett.120.151301",
    journal = {\PRL},
    volume = "120",
    number = "15",
    pages = "151301",
    year = "2018"
}

@article{ADMX:2019uok,
    author = "Braine, T. and others",
    collaboration = "ADMX",
    title = "{Extended Search for the Invisible Axion with the Axion Dark Matter Experiment}",
    eprint = "1910.08638",
    archivePrefix = "arXiv",
    primaryClass = "hep-ex",
    reportNumber = "FERMILAB-PUB-19-569-AD-AE-PPD",
    doi = "10.1103/PhysRevLett.124.101303",
    journal = {\PRL},
    volume = "124",
    number = "10",
    pages = "101303",
    year = "2020"
}

@article{ADMX:2021nhd,
    author = "Bartram, C. and others",
    collaboration = "ADMX",
    title = "{Search for Invisible Axion Dark Matter in the 3.3\textendash{}4.2\,\,\ensuremath{\mu}eV Mass Range}",
    eprint = "2110.06096",
    archivePrefix = "arXiv",
    primaryClass = "hep-ex",
    reportNumber = "FERMILAB-PUB-21-774-DI-PPD-SQMS",
    doi = "10.1103/PhysRevLett.127.261803",
    journal = {\PRL},
    volume = "127",
    number = "26",
    pages = "261803",
    year = "2021"
}

@article{Bartram:2024ovw,
    author = "Goodman, C. and others",
    collaboration = "ADMX",
    title = "{ADMX Axion Dark Matter Bounds around 3.3{\,}{\,}{\ensuremath{\mu}}eV with Dine-Fischler-Srednicki-Zhitnitsky Discovery Ability}",
    eprint = "2408.15227",
    archivePrefix = "arXiv",
    primaryClass = "hep-ex",
    reportNumber = "FERMILAB-PUB-24-0602-V",
    doi = "10.1103/PhysRevLett.134.111002",
    journal = {\PRL},
    volume = "134",
    number = "11",
    pages = "111002",
    year = "2025"
}

@article{ADMX:2025vom,
    author = "Carosi, G. and others",
    collaboration = "ADMX",
    title = "{Search for Axion Dark Matter from 1.1 to 1.3~GHz with ADMX}",
    eprint = "2504.07279",
    archivePrefix = "arXiv",
    primaryClass = "hep-ex",
    reportNumber = "FERMILAB-PUB-25-0298-PPD",
    doi = "10.1103/d7mg-6sqq",
    journal = {\PRL},
    volume = "135",
    number = "19",
    pages = "191001",
    year = "2025"
}

@article{ADMX:2018ogs,
    author = "Boutan, C. and others",
    collaboration = "ADMX",
    title = "{Piezoelectrically Tuned Multimode Cavity Search for Axion Dark Matter}",
    eprint = "1901.00920",
    archivePrefix = "arXiv",
    primaryClass = "hep-ex",
    reportNumber = "FERMILAB-PUB-18-702-AD-AE",
    doi = "10.1103/PhysRevLett.121.261302",
    journal = {\PRL},
    volume = "121",
    number = "26",
    pages = "261302",
    year = "2018"
}

@article{Bartram:2021ysp,
    author = "Bartram, C. and others",
    collaboration = "ADMX",
    title = "{Dark matter axion search using a Josephson Traveling wave parametric amplifier}",
    eprint = "2110.10262",
    archivePrefix = "arXiv",
    primaryClass = "hep-ex",
    reportNumber = "FERMILAB-PUB-22-615-PPD-SQMS",
    doi = "10.1063/5.0122907",
    journal = {\RSI},
    volume = "94",
    number = "4",
    pages = "044703",
    year = "2023"
}

@article{Crisosto:2019fcj,
    author = "Crisosto, N. and Sikivie, P. and Sullivan, N. S. and Tanner, D. B. and Yang, J. and Rybka, G.",
    title = "{ADMX SLIC: Results from a Superconducting $LC$ Circuit Investigating Cold Axions}",
    eprint = "1911.05772",
    archivePrefix = "arXiv",
    primaryClass = "astro-ph.CO",
    doi = "10.1103/PhysRevLett.124.241101",
    journal = {\PRL},
    volume = "124",
    number = "24",
    pages = "241101",
    year = "2020"
}

@article{Lee:2020cfj,
    author = "Lee, S. and Ahn, S. and Choi, J. and Ko, B. R. and Semertzidis, Y. K.",
    title = "{Axion Dark Matter Search around 6.7 $\mu$eV}",
    eprint = "2001.05102",
    archivePrefix = "arXiv",
    primaryClass = "hep-ex",
    doi = "10.1103/PhysRevLett.124.101802",
    journal = {\PRL},
    volume = "124",
    number = "10",
    pages = "101802",
    year = "2020"
}

@article{Jeong:2020cwz,
    author = "Jeong, Junu and Youn, SungWoo and Bae, Sungjae and Kim, Jihngeun and Seong, Taehyeon and Kim, Jihn E. and Semertzidis, Yannis K.",
    title = "{Search for Invisible Axion Dark Matter with a Multiple-Cell Haloscope}",
    eprint = "2008.10141",
    archivePrefix = "arXiv",
    primaryClass = "hep-ex",
    doi = "10.1103/PhysRevLett.125.221302",
    journal = {\PRL},
    volume = "125",
    number = "22",
    pages = "221302",
    year = "2020"
}

@article{CAPP:2020utb,
    author = "Kwon, Ohjoon and others",
    collaboration = "CAPP",
    title = "{First Results from an Axion Haloscope at CAPP around 10.7  $\mu$eV}",
    eprint = "2012.10764",
    archivePrefix = "arXiv",
    primaryClass = "hep-ex",
    doi = "10.1103/PhysRevLett.126.191802",
    journal = {\PRL},
    volume = "126",
    number = "19",
    pages = "191802",
    year = "2021"
}

@article{Yoon:2022gzp,
    author = "Yoon, Hojin and Ahn, Moohyun and Yang, Byeongsu and Lee, Youngjae and Kim, DongLak and Park, Heejun and Min, Byeonghun and Yoo, Jonghee",
    title = "{Axion haloscope using an 18~T high temperature superconducting magnet}",
    eprint = "2206.12271",
    archivePrefix = "arXiv",
    primaryClass = "hep-ex",
    doi = "10.1103/PhysRevD.106.092007",
    journal = {\PRD},
    volume = "106",
    number = "9",
    pages = "092007",
    year = "2022"
}

@article{Lee:2022mnc,
    author = "Lee, Youngjae and Yang, Byeongsu and Yoon, Hojin and Ahn, Moohyun and Park, Heejun and Min, Byeonghun and Kim, DongLak and Yoo, Jonghee",
    title = "{Searching for Invisible Axion Dark Matter with an 18~T Magnet Haloscope}",
    eprint = "2206.08845",
    archivePrefix = "arXiv",
    primaryClass = "hep-ex",
    doi = "10.1103/PhysRevLett.128.241805",
    journal = {\PRL},
    volume = "128",
    number = "24",
    pages = "241805",
    year = "2022"
}

@article{Kim:2022hmg,
    author = "Kim, Jinsu and others",
    title = "{Near-Quantum-Noise Axion Dark Matter Search at CAPP around 9.5{\,}{\,}{\ensuremath{\mu}}eV}",
    eprint = "2207.13597",
    archivePrefix = "arXiv",
    primaryClass = "hep-ex",
    doi = "10.1103/PhysRevLett.130.091602",
    journal = {\PRL},
    volume = "130",
    number = "9",
    pages = "091602",
    year = "2023"
}

@article{Yi:2022fmn,
    author = "Yi, Andrew K. and others",
    title = "{Axion Dark Matter Search around 4.55{\,}{\,}{\ensuremath{\mu}}eV with Dine-Fischler-Srednicki-Zhitnitskii Sensitivity}",
    eprint = "2210.10961",
    archivePrefix = "arXiv",
    primaryClass = "hep-ex",
    doi = "10.1103/PhysRevLett.130.071002",
    journal = {\PRL},
    volume = "130",
    number = "7",
    pages = "071002",
    year = "2023"
}

@article{Yang:2023yry,
    author = "Yang, Byeongsu and Yoon, Hojin and Ahn, Moohyun and Lee, Youngjae and Yoo, Jonghee",
    title = "{Extended Axion Dark Matter Search Using the CAPP18T Haloscope}",
    eprint = "2308.09077",
    archivePrefix = "arXiv",
    primaryClass = "hep-ex",
    doi = "10.1103/PhysRevLett.131.081801",
    journal = {\PRL},
    volume = "131",
    number = "8",
    pages = "081801",
    year = "2023"
}

@article{Kim:2023vpo,
    author = "Kim, Younggeun and others",
    title = "{Experimental Search for Invisible Dark Matter Axions around 22{\,}{\,}{\ensuremath{\mu}}eV}",
    eprint = "2312.11003",
    archivePrefix = "arXiv",
    primaryClass = "hep-ex",
    doi = "10.1103/PhysRevLett.133.051802",
    journal = {\PRL},
    volume = "133",
    number = "5",
    pages = "051802",
    year = "2024"
}

@article{CAPP:2024dtx,
    author = "Ahn, Saebyeok and others",
    collaboration = "CAPP",
    title = "{Extensive Search for Axion Dark Matter over 1~GHz with CAPP{\textquoteright}S Main Axion Experiment}",
    eprint = "2402.12892",
    archivePrefix = "arXiv",
    primaryClass = "hep-ex",
    doi = "10.1103/PhysRevX.14.031023",
    journal = "\PRX",
    volume = "14",
    number = "3",
    pages = "031023",
    year = "2024"
}

@article{Bae:2024kmy,
    author = "Bae, Sungjae and Jeong, Junu and Kim, Younggeun and Youn, SungWoo and Park, Heejun and Seong, Taehyeon and Oh, Seongjeong and Semertzidis, Yannis K.",
    title = "{Search for Dark Matter Axions with Tunable TM020 Mode}",
    eprint = "2403.13390",
    archivePrefix = "arXiv",
    primaryClass = "hep-ex",
    doi = "10.1103/PhysRevLett.133.211803",
    journal = {\PRL},
    volume = "133",
    number = "21",
    pages = "211803",
    year = "2024"
}

@article{Adair:2022rtw,
    author = "Adair, C. M. and others",
    title = "{Search for Dark Matter Axions with CAST-CAPP}",
    eprint = "2211.02902",
    archivePrefix = "arXiv",
    primaryClass = "hep-ex",
    doi = "10.1038/s41467-022-33913-6",
    journal = {\NatC},
    volume = "13",
    number = "1",
    pages = "6180",
    year = "2022"
}

@article{Oshima:2023csb,
    author = "Oshima, Yuka and Fujimoto, Hiroki and Kume, Jun'ya and Morisaki, Soichiro and Nagano, Koji and Fujita, Tomohiro and Obata, Ippei and Nishizawa, Atsushi and Michimura, Yuta and Ando, Masaki",
    title = "{First results of axion dark matter search with DANCE}",
    eprint = "2303.03594",
    archivePrefix = "arXiv",
    primaryClass = "hep-ex",
    reportNumber = "RESCEU-4/23",
    doi = "10.1103/PhysRevD.108.072005",
    journal = {\PRD},
    volume = "108",
    number = "7",
    pages = "072005",
    year = "2023"
}

@article{Devlin:2021fpq,
    author = "Devlin, Jack A. and others",
    title = "{Constraints on the Coupling between Axionlike Dark Matter and Photons Using an Antiproton Superconducting Tuned Detection Circuit in a Cryogenic Penning Trap}",
    eprint = "2101.11290",
    archivePrefix = "arXiv",
    primaryClass = "astro-ph.CO",
    doi = "10.1103/PhysRevLett.126.041301",
    journal = {\PRL},
    volume = "126",
    number = "4",
    pages = "041301",
    year = "2021"
}

@article{Hoshino:2025fiz,
    author = "Hoshino, Gabe and others",
    collaboration = "GigaBREAD",
    title = "{First Axionlike Particle Results from a Broadband Search for Wavelike Dark Matter in the 44 to 52{\,}{\,}{\ensuremath{\mu}}eV Range with a Coaxial Dish Antenna}",
    eprint = "2501.17119",
    archivePrefix = "arXiv",
    primaryClass = "hep-ex",
    reportNumber = "FERMILAB-PUB-25-0076-PPD",
    doi = "10.1103/PhysRevLett.134.171002",
    journal = {\PRL},
    volume = "134",
    number = "17",
    pages = "171002",
    year = "2025"
}

@article{Li_2020,
doi = {10.1088/1742-6596/1468/1/012062},
url = {https://doi.org/10.1088/1742-6596/1468/1/012062},
year = {2020},
month = {feb},
publisher = {IOP Publishing},
volume = {1468},
number = {1},
pages = {012062},
author = {Li, Xiaoyue and for the MADMAX Collaboration},
title = {MADMAX: A Dielectric Haloscope Experiment},
journal = {Journal of Physics: Conference Series}
}

@article{Millar_2023,
  title = {Searching for dark matter with plasma haloscopes},
  author = {Millar, Alexander J. and Anlage, Steven M. and Balafendiev, Rustam and Belov, Pavel and van Bibber, Karl and Conrad, Jan and Demarteau, Marcel and Droster, Alexander and Dunne, Katherine and Rosso, Andrea Gallo and Gudmundsson, Jon E. and Jackson, Heather and Kaur, Gagandeep and Klaesson, Tove and Kowitt, Nolan and Lawson, Matthew and Leder, Alexander and Miyazaki, Akira and Morampudi, Sid and Peiris, Hiranya V. and R\o{}ising, Henrik S. and Singh, Gaganpreet and Sun, Dajie and Thomas, Jacob H. and Wilczek, Frank and Withington, Stafford and Wooten, Mackenzie and Dilling, Jens and Febbraro, Michael and Knirck, Stefan and Marvinney, Claire},
  collaboration = {Endorsers},
  journal = {Phys. Rev. D},
  volume = {107},
  issue = {5},
  pages = {055013},
  numpages = {26},
  year = {2023},
  month = {Mar},
  publisher = {American Physical Society},
  doi = {10.1103/PhysRevD.107.055013},
  url = {https://link.aps.org/doi/10.1103/PhysRevD.107.055013}
}

@article{Grenet:2021vbb,
    author = "Grenet, Thierry and Ballou, Rafik and Basto, Quentin and Martineau, Killian and Perrier, Pierre and Pugnat, Pierre and Quevillon, J\'er\'emie and Roch, Nicolas and Smith, Christopher",
    title = "{The Grenoble Axion Haloscope platform (GrAHal): development plan and first results}",
    eprint = "2110.14406",
    archivePrefix = "arXiv",
    primaryClass = "hep-ex",
    month = "10",
    year = "2021",
    journal=""
}

@article{Brubaker:2016ktl,
    author = "Brubaker, B. M. and others",
    title = "{First results from a microwave cavity axion search at 24 $\mu$eV}",
    eprint = "1610.02580",
    archivePrefix = "arXiv",
    primaryClass = "astro-ph.CO",
    doi = "10.1103/PhysRevLett.118.061302",
    journal = {\PRL},
    volume = "118",
    number = "6",
    pages = "061302",
    year = "2017"
}

@article{HAYSTAC:2018rwy,
    author = "Zhong, L. and others",
    collaboration = "HAYSTAC",
    title = "{Results from phase 1 of the HAYSTAC microwave cavity axion experiment}",
    eprint = "1803.03690",
    archivePrefix = "arXiv",
    primaryClass = "hep-ex",
    doi = "10.1103/PhysRevD.97.092001",
    journal = {\PRD},
    volume = "97",
    number = "9",
    pages = "092001",
    year = "2018"
}

@article{HAYSTAC:2020kwv,
    author = "Backes, K. M. and others",
    collaboration = "HAYSTAC",
    title = "{A quantum-enhanced search for dark matter axions}",
    eprint = "2008.01853",
    archivePrefix = "arXiv",
    primaryClass = "quant-ph",
    doi = "10.1038/s41586-021-03226-7",
    journal = "Nature",
    volume = "590",
    number = "7845",
    pages = "238--242",
    year = "2021"
}

@article{HAYSTAC:2023cam,
    author = "Jewell, M. J. and others",
    collaboration = "HAYSTAC",
    title = "{New results from HAYSTAC{\textquoteright}s phase II operation with a squeezed state receiver}",
    eprint = "2301.09721",
    archivePrefix = "arXiv",
    primaryClass = "hep-ex",
    doi = "10.1103/PhysRevD.107.072007",
    journal = {\PRD},
    volume = "107",
    number = "7",
    pages = "072007",
    year = "2023"
}

@article{HAYSTAC:2024jch,
    author = "Bai, Xiran and others",
    collaboration = "HAYSTAC",
    title = "{Dark Matter Axion Search with HAYSTAC Phase II}",
    eprint = "2409.08998",
    archivePrefix = "arXiv",
    primaryClass = "hep-ex",
    doi = "10.1103/PhysRevLett.134.151006",
    journal = {\PRL},
    volume = "134",
    number = "15",
    pages = "151006",
    year = "2025"
}

@article{Heinze:2023nfb,
    author = "Heinze, Joscha and Gill, Alex and Dmitriev, Artemiy and Smetana, Jiri and Yan, Tianliang and Boyer, Vincent and Martynov, Denis and Evans, Matthew",
    title = "{First Results of the Laser-Interferometric Detector for Axions (LIDA)}",
    eprint = "2307.01365",
    archivePrefix = "arXiv",
    primaryClass = "astro-ph.CO",
    doi = "10.1103/PhysRevLett.132.191002",
    journal = {\PRL},
    volume = "132",
    number = "19",
    pages = "191002",
    year = "2024"
}

@article{Garcia:2024xzc,
    author = "Garcia, B. Ary dos Santos and others",
    collaboration = "MADMAX",
    title = "{First Search for Axion Dark Matter with a MADMAX Prototype}",
    eprint = "2409.11777",
    archivePrefix = "arXiv",
    primaryClass = "hep-ex",
    reportNumber = "FERMILAB-PUB-24-0095-PPD",
    doi = "10.1103/c749-419q",
    journal = {\PRL},
    volume = "135",
    number = "4",
    pages = "041001",
    year = "2025"
}

@article{McAllister:2017lkb,
    author = "McAllister, Ben T. and Flower, Graeme and Ivanov, Eugene N. and Goryachev, Maxim and Bourhill, Jeremy and Tobar, Michael E.",
    title = "{The ORGAN Experiment: An axion haloscope above 15 GHz}",
    eprint = "1706.00209",
    archivePrefix = "arXiv",
    primaryClass = "physics.ins-det",
    doi = "10.1016/j.dark.2017.09.010",
    journal = {\PDU},
    volume = "18",
    pages = "67--72",
    year = "2017"
}

@article{Quiskamp:2022pks,
    author = "Quiskamp, Aaron P. and McAllister, Ben T. and Altin, Paul and Ivanov, Eugene N. and Goryachev, Maxim and Tobar, Michael E.",
    title = "{Direct search for dark matter axions excluding ALP cogenesis in the 63- to 67-\ensuremath{\mu}eV range with the ORGAN experiment}",
    eprint = "2203.12152",
    archivePrefix = "arXiv",
    primaryClass = "hep-ex",
    doi = "10.1126/sciadv.abq3765",
    journal = {\SA},
    volume = "8",
    number = "27",
    pages = "abq3765",
    year = "2022"
}

@article{Quiskamp:2023ehr,
    author = "Quiskamp, Aaron and McAllister, Ben T. and Altin, Paul and Ivanov, Eugene N. and Goryachev, Maxim and Tobar, Michael E.",
    title = "{Exclusion of Axionlike-Particle Cogenesis Dark Matter in a Mass Window above 100{\,}{\,}{\ensuremath{\mu}}eV}",
    eprint = "2310.00904",
    archivePrefix = "arXiv",
    primaryClass = "hep-ex",
    doi = "10.1103/PhysRevLett.132.031601",
    journal = {\PRL},
    volume = "132",
    number = "3",
    pages = "031601",
    year = "2024"
}

@article{Quiskamp:2024oet,
    author = "Quiskamp, Aaron P. and Flower, Graeme R. and Samuels, Steven and McAllister, Ben T. and Altin, Paul and Ivanov, Eugene N. and Goryachev, Maxim and Tobar, Michael E.",
    title = "{Near-quantum-limited axion dark matter search with the ORGAN experiment around 26{\,}{\,}{\ensuremath{\mu}}eV}",
    eprint = "2407.18586",
    archivePrefix = "arXiv",
    primaryClass = "hep-ex",
    doi = "10.1103/PhysRevD.111.095007",
    journal = {\PRD},
    volume = "111",
    number = "9",
    pages = "095007",
    year = "2025"
}

@article{Alesini:2019ajt,
    author = "Alesini, D. and others",
    title = "{Galactic axions search with a superconducting resonant cavity}",
    eprint = "1903.06547",
    archivePrefix = "arXiv",
    primaryClass = "physics.ins-det",
    doi = "10.1103/PhysRevD.99.101101",
    journal = {\PRD},
    volume = "99",
    number = "10",
    pages = "101101",
    year = "2019"
}

@article{Alesini:2020vny,
    author = "Alesini, D. and others",
    title = "{Search for invisible axion dark matter of mass m$_a=43~\mu$eV with the QUAX--$a\gamma$ experiment}",
    eprint = "2012.09498",
    archivePrefix = "arXiv",
    primaryClass = "hep-ex",
    doi = "10.1103/PhysRevD.103.102004",
    journal = {\PRD},
    volume = "103",
    number = "10",
    pages = "102004",
    year = "2021"
}

@article{Alesini:2022lnp,
    author = "Alesini, D. and others",
    title = "{Search for Galactic axions with a high-Q dielectric cavity}",
    eprint = "2208.12670",
    archivePrefix = "arXiv",
    primaryClass = "hep-ex",
    doi = "10.1103/PhysRevD.106.052007",
    journal = {\PRD},
    volume = "106",
    number = "5",
    pages = "052007",
    year = "2022"
}

@article{QUAX:2023gop,
    author = "Di Vora, R. and others",
    collaboration = "QUAX",
    title = "{Search for galactic axions with a traveling wave parametric amplifier}",
    eprint = "2304.07505",
    archivePrefix = "arXiv",
    primaryClass = "hep-ex",
    reportNumber = "FERMILAB-PUB-23-229-SQMS-V",
    doi = "10.1103/PhysRevD.108.062005",
    journal = {\PRD},
    volume = "108",
    number = "6",
    pages = "062005",
    year = "2023"
}

@article{QUAX:2024fut,
    author = "Rettaroli, A. and others",
    collaboration = "QUAX",
    title = "{Search for axion dark matter with the QUAX\textendash{}LNF tunable haloscope}",
    eprint = "2402.19063",
    archivePrefix = "arXiv",
    primaryClass = "physics.ins-det",
    reportNumber = "FERMILAB-PUB-24-0511-SQMS-V",
    doi = "10.1103/PhysRevD.110.022008",
    journal = {\PRD},
    volume = "110",
    number = "2",
    pages = "022008",
    year = "2024"
}

@article{CAST:2020rlf,
    author = "Melc\'on, A. \'Alvarez and others",
    collaboration = "CAST",
    title = "{First results of the CAST-RADES haloscope search for axions at 34.67 $\mu$eV}",
    eprint = "2104.13798",
    archivePrefix = "arXiv",
    primaryClass = "hep-ex",
    reportNumber = "CERN-EP-2021-070",
    doi = "10.1007/JHEP10(2021)075",
    journal = {\JHEP},
    volume = "21",
    pages = "075",
    year = "2020"
}

@article{Ahyoune:2024klt,
    author = "Ahyoune, S. and others",
    title = "{RADES axion search results with a high-temperature superconducting cavity in an 11.7 T magnet}",
    eprint = "2403.07790",
    archivePrefix = "arXiv",
    primaryClass = "hep-ex",
    reportNumber = "CERN-EP-2024-076, MPP-2024-55",
    doi = "10.1007/JHEP04(2025)113",
    journal = {\JHEP},
    volume = "04",
    pages = "113",
    year = "2025"
}

@article{DePanfilis,
  title = {Limits on the abundance and coupling of cosmic axions at 4.5$<{m}_{a}<$5.0 \ensuremath{\mu}eV},
  author = {DePanfilis, S. and Melissinos, A. C. and Moskowitz, B. E. and Rogers, J. T. and Semertzidis, Y. K. and Wuensch, W. U. and Halama, H. J. and Prodell, A. G. and Fowler, W. B. and Nezrick, F. A.},
  journal = {\PRL},
  volume = {59},
  issue = {7},
  pages = {839--842},
  numpages = {0},
  year = {1987},
  month = {Aug},
  publisher = {American Physical Society},
  doi = {10.1103/PhysRevLett.59.839},
  url = {https://link.aps.org/doi/10.1103/PhysRevLett.59.839}
}

@article{Wuensch:1989sa,
    author = "Wuensch, Walter and De Panfilis-Wuensch, S. and Semertzidis, Y. K. and Rogers, J. T. and Melissinos, A. C. and Halama, H. J. and Moskowitz, B. E. and Prodell, A. G. and Fowler, W. B. and Nezrick, F. A.",
    title = "{Results of a Laboratory Search for Cosmic Axions and Other Weakly Coupled Light Particles}",
    reportNumber = "FERMILAB-PUB-89-185-E, BNL-43010",
    doi = "10.1103/PhysRevD.40.3153",
    journal = {\PRD},
    volume = "40",
    pages = "3153",
    year = "1989"
}

@article{Gramolin:2020ict,
    author = "Gramolin, Alexander V. and Aybas, Deniz and Johnson, Dorian and Adam, Janos and Sushkov, Alexander O.",
    title = "{Search for axion-like dark matter with ferromagnets}",
    eprint = "2003.03348",
    archivePrefix = "arXiv",
    primaryClass = "hep-ex",
    doi = "10.1038/s41567-020-1006-6",
    journal = {\NatP},
    volume = "17",
    number = "1",
    pages = "79--84",
    year = "2021"
}

@article{TASEH:2022vvu,
    author = "Chang, Hsin and others",
    collaboration = "TASEH",
    title = "{First Results from the Taiwan Axion Search Experiment with a Haloscope at 19.6\,\,\ensuremath{\mu}eV}",
    eprint = "2205.05574",
    archivePrefix = "arXiv",
    primaryClass = "hep-ex",
    doi = "10.1103/PhysRevLett.129.111802",
    journal = {\PRL},
    volume = "129",
    number = "11",
    pages = "111802",
    year = "2022"
}

@article{Arza:2021ekq,
    author = "Arza, Ariel and Fedderke, Michael A. and Graham, Peter W. and Kimball, Derek F. Jackson and Kalia, Saarik",
    title = "{Earth as a transducer for axion dark-matter detection}",
    eprint = "2112.09620",
    archivePrefix = "arXiv",
    primaryClass = "hep-ph",
    doi = "10.1103/PhysRevD.105.095007",
    journal = {\PRD},
    volume = "105",
    number = "9",
    pages = "095007",
    year = "2022"
}

@article{Friel:2024shg,
    author = "Friel, Matt and Gjerloev, Jesper W. and Kalia, Saarik and Zamora, Alvaro",
    title = "{Search for ultralight dark matter in the SuperMAG high-fidelity dataset}",
    eprint = "2408.16045",
    archivePrefix = "arXiv",
    primaryClass = "hep-ph",
    reportNumber = "FERMILAB-PUB-24-0546-SQMS",
    doi = "10.1103/PhysRevD.110.115036",
    journal = {\PRD},
    volume = "110",
    number = "11",
    pages = "115036",
    year = "2024"
}

@article{Nishizawa:2025xka,
    author = "Nishizawa, Atsushi and Taruya, Atsushi and Himemoto, Yoshiaki",
    title = "{Axion dark matter search from terrestrial magnetic fields at extremely low frequencies}",
    eprint = "2504.07559",
    archivePrefix = "arXiv",
    primaryClass = "hep-ph",
    reportNumber = "YITP-25-55",
    month = "4",
    year = "2025",
    journal = ""
}

@article{Hagmann,
  title = {Results from a search for cosmic axions},
  author = {Hagmann, C. and Sikivie, P. and Sullivan, N. S. and Tanner, D. B.},
  journal = {\PRD},
  volume = {42},
  issue = {4},
  pages = {1297--1300},
  numpages = {0},
  year = {1990},
  month = {Aug},
  publisher = {American Physical Society},
  doi = {10.1103/PhysRevD.42.1297},
  url = {https://link.aps.org/doi/10.1103/PhysRevD.42.1297}
}

@article{Hagmann:1996qd,
    author = "Hagmann, C. and others",
    editor = "Cline, D. B.",
    title = "{First results from a second generation galactic axion experiment}",
    eprint = "astro-ph/9607022",
    archivePrefix = "arXiv",
    reportNumber = "UCRL-JC-124162",
    doi = "10.1016/S0920-5632(96)00516-6",
    journal = {\NPBPS},
    volume = "51",
    pages = "209--212",
    year = "1996"
}

@article{Thomson:2019aht,
    author = "Thomson, Catriona A. and McAllister, Ben T. and Goryachev, Maxim and Ivanov, Eugene N. and Tobar, Michael E.",
    title = "{Upconversion Loop Oscillator Axion Detection Experiment: A Precision Frequency Interferometric Axion Dark Matter Search with a Cylindrical Microwave Cavity}",
    eprint = "1912.07751",
    archivePrefix = "arXiv",
    primaryClass = "hep-ex",
    doi = "10.1103/PhysRevLett.127.019901",
    journal = {\PRL},
    volume = "126",
    number = "8",
    pages = "081803",
    year = "2021",
    note = "[Erratum: \href{https://doi.org/10.1103/PhysRevLett.127.019901}{\PRL 127, 019901 (2021)}]"
}

@article{Thomson:2023moc,
    author = "Thomson, Catriona A. and Goryachev, Maxim and McAllister, Ben T. and Ivanov, Eugene N. and Altin, Paul and Tobar, Michael E.",
    title = "{Searching for low-mass axions using resonant upconversion}",
    eprint = "2301.06778",
    archivePrefix = "arXiv",
    primaryClass = "hep-ex",
    doi = "10.1103/PhysRevD.107.112003",
    journal = {\PRD},
    volume = "107",
    number = "11",
    pages = "112003",
    year = "2023"
}

@article{Ehret:2010mh,
    author = "Ehret, Klaus and others",
    title = "{New ALPS Results on Hidden-Sector Lightweights}",
    eprint = "1004.1313",
    archivePrefix = "arXiv",
    primaryClass = "hep-ex",
    reportNumber = "DESY-10-030, MPP-2010-27",
    doi = "10.1016/j.physletb.2010.04.066",
    journal = {\PLB},
    volume = "689",
    pages = "149--155",
    year = "2010"
}

@article{CAST:2007jps,
    author = "Andriamonje, S. and others",
    collaboration = "CAST",
    title = "{An improved limit on the axion-photon coupling from the CAST experiment}",
    eprint = "hep-ex/0702006",
    archivePrefix = "arXiv",
    doi = "10.1088/1475-7516/2007/04/010",
    journal = {\JCAP},
    volume = "04",
    pages = "010",
    year = "2007"
}

@article{CAST:2017uph,
    author = "Anastassopoulos, V. and others",
    collaboration = "CAST",
    title = "{New CAST Limit on the Axion-Photon Interaction}",
    eprint = "1705.02290",
    archivePrefix = "arXiv",
    primaryClass = "hep-ex",
    doi = "10.1038/nphys4109",
    journal = {\NatP},
    volume = "13",
    pages = "584--590",
    year = "2017"
}

@article{CAST:2024eil,
    author = {Altenm{\"u}ller, K. and others},
    collaboration = "CAST",
    title = "{New Upper Limit on the Axion-Photon Coupling with an Extended CAST Run with a Xe-Based Micromegas Detector}",
    eprint = "2406.16840",
    archivePrefix = "arXiv",
    primaryClass = "hep-ex",
    doi = "10.1103/PhysRevLett.133.221005",
    journal = {\PRL},
    volume = "133",
    number = "22",
    pages = "221005",
    year = "2024"
}

@article{Betz:2013dza,
    author = "Betz, M. and Caspers, F. and Gasior, M. and Thumm, M. and Rieger, S. W.",
    title = "{First results of the CERN Resonant Weakly Interacting sub-eV Particle Search (CROWS)}",
    eprint = "1310.8098",
    archivePrefix = "arXiv",
    primaryClass = "physics.ins-det",
    doi = "10.1103/PhysRevD.88.075014",
    journal = {\PRD},
    volume = "88",
    number = "7",
    pages = "075014",
    year = "2013"
}

@article{OSQAR:2015qdv,
    author = "Ballou, R. and others",
    collaboration = "OSQAR",
    title = "{New exclusion limits on scalar and pseudoscalar axionlike particles from light shining through a wall}",
    eprint = "1506.08082",
    archivePrefix = "arXiv",
    primaryClass = "hep-ex",
    doi = "10.1103/PhysRevD.92.092002",
    journal = {\PRD},
    volume = "92",
    number = "9",
    pages = "092002",
    year = "2015"
}

@article{DellaValle:2015xxa,
    author = "Della Valle, Federico and Ejlli, Aldo and Gastaldi, Ugo and Messineo, Giuseppe and Milotti, Edoardo and Pengo, Ruggero and Ruoso, Giuseppe and Zavattini, Guido",
    title = "{The PVLAS experiment: measuring vacuum magnetic birefringence and dichroism with a birefringent Fabry\textendash{}Perot cavity}",
    eprint = "1510.08052",
    archivePrefix = "arXiv",
    primaryClass = "physics.optics",
    doi = "10.1140/epjc/s10052-015-3869-8",
    journal = {\EPJC},
    volume = "76",
    number = "1",
    pages = "24",
    year = "2016"
}

@article{SAPPHIRES:2021vkz,
    author = "Homma, Kensuke and others",
    collaboration = "SAPPHIRES",
    title = "{Search for sub-eV axion-like resonance states via stimulated quasi-parallel laser collisions with the parameterization including fully asymmetric collisional geometry}",
    eprint = "2105.01224",
    archivePrefix = "arXiv",
    primaryClass = "hep-ex",
    doi = "10.1007/JHEP12(2021)108",
    journal = {\JHEP},
    volume = "12",
    pages = "108",
    year = "2021"
}

@article{Chan:2021gjl,
    author = "Chan, Man Ho",
    title = "{Constraining the axion{\textendash}photon coupling using radio data of the Bullet cluster}",
    eprint = "2109.11734",
    archivePrefix = "arXiv",
    primaryClass = "astro-ph.CO",
    doi = "10.1038/s41598-021-99495-3",
    journal = "Sci. Rep.",
    volume = "11",
    number = "1",
    pages = "20087",
    year = "2021"
}

\end{document}